\RequirePackage{fix-cm}
\documentclass[twocolumn,epjc3]{svjour3}
\RequirePackage[colorlinks,citecolor=blue,urlcolor=blue,linkcolor=blue]{hyperref}
\RequirePackage{cite}
\RequirePackage{graphicx}
\RequirePackage{flushend}
\RequirePackage{amssymb}
\RequirePackage{amsmath}
\begin{document}

\title{
Radially and azimuthally excited states of a soliton system of vortex and Q-ball
}
\author{A.Yu.~Loginov\thanksref{addr1,e1}
        \and
        V.V.~Gauzshtein\thanksref{addr2,e2} 
}
\thankstext{e1}{e-mail: a.yu.loginov@tusur.ru}
\thankstext{e2}{e-mail: gauzshtein@tpu.ru}

\institute{Tomsk State University of Control Systems and Radioelectronics, 634050 Tomsk, Russia \label{addr1}
           \and
           Tomsk Polytechnic University, 634050 Tomsk, Russia \label{addr2}
}
\date{Received: date / Accepted: date}
\maketitle

\begin{abstract}
In the present paper, we continue  to  study the two-dimensional soliton system
that is composed of vortex and Q-ball  components  interacting  with each other
through an Abelian gauge field.
This vortex-Q-ball system is electrically neutral  as  a whole, nevertheless it
possesses a nonzero electric field.
Moreover, the vortex-Q-ball system has  a quantized magnetic flux and a nonzero
angular momentum,  and  combines  properties  of topological and nontopological
solitons.
We investigate radially  and  azimuthally  excited  states of the vortex-Q-ball
system along with  the  unexcited  vortex-Q-ball  system at different values of
gauge coupling constants.
We also ascertain the behaviour of  the vortex-Q-ball system in several extreme
regimes, including thin-wall and thick-wall regimes.

\end{abstract}

\section{Introduction} \label{sec:I}

It is known that $(1+1)$-dimensional gauge models and $(2+1)$-dimensional gauge
models without the Chern-Simons term do not allow the existence of electrically
charged solitons because  any  electrically  charged  compact  object will have
infinite energy in these models.
The reason for  this  is  simple: at large  distances,  the electric field of a
one-dimensional object  does  not  depend  on   the   distance  and  that  of a
two-dimensional object is inversely proportional to the distance.
As a result,  with  increasing  distance, the  energy  of  the  electric  field
diverges  linearly  in  the  one-dimensional  case  and  logarithmically in the
two-dimensional case.

The  electrically  charged  solitons  appear only  in $(3+1)$-dimensional gauge
models (e.g. the three-dimensional electrically charged  dyon  \cite{julia_zee}
or Q-ball \cite{rosen,klee,anag,ardoz_2009,tamaki_2014,gulamov_2014}).
Note, however, that the Chern-Simons term  can  be added to the Lagrangian of a
$(2+1)$-dimensional gauge model.
Moreover, $(2+1)$-dimensional gauge  models may be pure Chern-Simons, and hence
have no Maxwell gauge term.
Such $(2+1)$-dimensional gauge  models may  admit the existence of electrically
charged solitons.
Indeed, electrically charged  vortices were found in both the pure Chern-Simons
\cite{hong,jw1,jw2,bazeia_1991,ghosh}     and      the     Maxwell-Chern-Simons
\cite{paul,khare_rao_227,khare_255,loginov_plb_784} gauge models.
In addition, one-dimensional  domain  walls  may  exist  in  Chern-Simons gauge
models \cite{dsantos,losano}.
These Chern-Simons  domain  walls  possess  finite  linear densities of energy,
magnetic flux, and electric charge.

Thus, in $(1+1)$ and $(2+1)$-dimensional pure Maxwell  gauge  models,  solitons
should be electrically neutral.
The  neutrality,  however,  does  not  mean  the  absence  of  electric  field.
In Refs. \cite{loginov_plb_777,loginov_epj_79}, one and two-dimensional soliton
systems composed of  topological and  nontopological components were described.
The components interact with each  other  through  an  Abelian  gauge field and
possess opposite electric  charges,  so  the  soliton  systems are neutral as a
whole.
Despite electrical neutrality, these soliton systems possess a nonzero electric
field that tends to zero exponentially at spatial infinity, resulting in finite
electrostatic energy.

The characteristic  feature  of  nontopological  solitons  is  the  presence of
radially           and            azimuthally         excited            states
\cite{FLS_1976,volkov_prd_66,kunz_prd_72,mai_prd_86,loginov_prd_102}.
The Q-ball components of  compound  soliton  systems  may  also  be radially or
azimuthally excited.
The corresponding excited compound soliton systems  will have some new features
compared to unexcited systems.
In the present paper, we study radially and azimuthally  excited  states of the
vortex-Q-ball soliton system described in \cite{loginov_plb_777}.
We also study the  unexcited  vortex-Q-ball  system  using  different values of
gauge coupling constants.

The paper is structured as follows.
In Sect.~\ref{sec:II},  we  describe  the Lagrangian, the symmetries, the field
equations,  and  the  energy-momentum   tensor   of   the   gauge  model  under
consideration.
In Sect.~\ref{sec:III}, we  list some  properties  of the vortex-Q-ball system;
among them,  the  basic  differential  relation,  the  asymptotic  behaviour of
fields at small and  large  distances, some  properties of the gauge potential,
the  virial  relation,  and  the  Laue  condition for the vortex-Q-ball system.
In Sect.~\ref{sec:IV},  we study  properties  of  the  vortex-Q-ball  system at
extreme values of parameters.
In Sect.~\ref{sec:V},  we  present  and  discuss the numerical results for  the
unexcited vortex-Q-ball system at different values of gauge coupling constants,
the  radially  excited  vortex-Q-ball  system,   and  the  azimuthally  excited
vortex-Q-ball system.
In all three cases, we present dependences of the  vortex-Q-ball  energy on the
phase frequency  and  on  the  Noether  charge  along  with radial  dependences
of the vortex-Q-ball ansatz functions.

Throughout the paper, we use the natural units $\hbar = c = 1$.

\section{The gauge model}                                        \label{sec:II}

The Lagrangian density  of the $\left(2+1\right)$-dimensional gauge model under
consideration has the form
\begin{eqnarray}
\mathcal{L} &=&-\frac{1}{4}F_{\mu \nu }F^{\mu \nu }
+ \left( D_{\mu }\phi \right) ^{\ast }D^{\mu }\phi - V\left( \left\vert \phi
\right\vert \right) \nonumber
\\
&&+\left( D_{\mu }\chi \right)^{\ast }D^{\mu }\chi - U\left( \left\vert \chi
\right\vert \right).                                               \label{II:1}
\end{eqnarray}
The model describes the  two  complex  scalar  fields  $\phi$  and  $\chi$ that
minimally interact with the Abelian gauge field $A_{\mu}$ through the covariant
derivatives
\begin{equation}
D_{\mu }\phi = \partial _{\mu }\phi + ieA_{\mu }\phi,\quad
D_{\mu }\chi = \partial _{\mu }\chi + iqA_{\mu }\chi.              \label{II:2}
\end{equation}
The scalar fields $\phi$ and $\chi$ are self-interacting ones.
The  self-interaction of $\phi$  and  $\chi$  is  described  by the fourth- and
sixth-order potentials, respectively
\begin{subequations} \label{II:3}
\begin{eqnarray}
V\left( \left\vert \phi \right\vert \right) & = & \frac{\lambda }{2}\left(
\left\vert \phi \right\vert ^{2}-v^{2}\right) ^{2},               \label{II:3a}
\\
U\left( \left\vert \chi \right\vert \right) & = & m^{2}\left\vert \chi
\right\vert ^{2}-\frac{g}{2}\left\vert \chi \right\vert ^{4}+\frac{h}{3}
\left\vert \chi \right\vert^{6},                                  \label{II:3b}
\end{eqnarray}
\end{subequations}
where $\left\vert \phi \right\vert^{2} = \phi^{\ast}\phi $ and $\left\vert \chi
\right\vert^{2} = \chi^{\ast}\chi$.
In Eqs.~(\ref{II:3a})  and  (\ref{II:3b}), $\lambda$, $g$, and $h$ are positive
self-interaction  constants, $m$ is the mass of the scalar $\chi$-particle, and
$v$ is the vacuum average of the  amplitude of the complex scalar field $\phi$.
From Eq.~(\ref{II:3a}) it follows that  the  potential $V\left( \left\vert \phi
\right\vert \right)$ possesses  the  continuous  family  of minima lying on the
circle $\left\vert \phi \right\vert = v$.
At the same time,  we suppose that the potential $U\left(\left\vert \chi \right
\vert \right)$ has a global isolated minimum at $\chi = 0$.
For this to hold, the parameters of $U\left(\left\vert \chi\right \vert\right)$
must satisfy the inequality $3g^{2} < 16m^{2}h$.

The  invariance  of  the  Lagrangian  density (\ref{II:1})  under  local  gauge
transformations
\begin{eqnarray}
\phi \left( x\right) &\rightarrow &\phi ^{\prime }\left( x\right) = \exp
\left( -i e\Lambda \left( x\right) \right) \phi \left( x\right),    \nonumber
\\
\chi \left( x\right) &\rightarrow &\chi ^{\prime }\left( x\right) = \exp
\left( -iq\Lambda \left( x\right) \right) \chi \left( x\right),     \nonumber
\\
A_{\mu }\left( x\right) & \rightarrow & A_{\mu }^{\prime}\left(x\right)
=A_{\mu }\left( x\right)+\partial_{\mu}\Lambda \left(x\right)      \label{II:4}
\end{eqnarray}
and the electrical neutrality of the Abelian  gauge field $A_{\mu}$ lead to the
invariance of the model under the two independent  global gauge transformations
\begin{eqnarray}
\phi \left( x\right) &\rightarrow &\phi ^{\prime }\left( x\right) = \exp
\left( -i\alpha \right) \phi \left( x\right),                         \nonumber
\\
\chi \left( x\right) &\rightarrow &\chi ^{\prime }\left( x\right) = \exp
\left( -i\beta \right) \chi \left( x\right).                       \label{II:5}
\end{eqnarray}
The  invariance  of  the   Lagrangian   density   under  global transformations
(\ref{II:5}) results in the existence of  the  two  conserved  Noether currents
\begin{eqnarray}
j_{\phi }^{\mu } &=&i\left[ \phi ^{\ast }D^{\mu }\phi -\left( D^{\mu }\phi
\right) ^{\ast }\phi \right],                                         \nonumber
\\
j_{\chi }^{\mu } &=&i\left[ \chi ^{\ast }D^{\mu }\chi -\left( D^{\mu }\chi
\right) ^{\ast }\chi \right].                                      \label{II:6}
\end{eqnarray}
The field equations for the model have the form
\begin{eqnarray}
D_{\mu }D^{\mu }\phi +\lambda \left( \left\vert \phi \right\vert
^{2}-v^{2}\right) \phi  & = &0,                                   \label{II:7a}
\\
D_{\mu }D^{\mu }\chi +m^{2}\chi -g\left\vert \chi \right\vert ^{2}\chi
+h\left\vert \chi \right\vert ^{4}\chi  & = &0,                   \label{II:7b}
\\
\partial _{\mu }F^{\mu \nu } & = &j^{\nu},                        \label{II:7c}
\end{eqnarray}
where the electromagnetic current $j^{\nu}$ is
\begin{eqnarray}
j^{\nu } &=&ej_{\phi }^{\mu }+qj_{\chi }^{\mu }                    \label{II:8}
\\
&=&ie\phi ^{\ast }\overleftrightarrow{\partial ^{\nu }}\phi -2e^{2}A^{\nu
}\phi^{\ast}\phi + i q \chi^{\ast}\overleftrightarrow{\partial^{\nu}}\chi
- 2 q^{2} A^{\nu}\chi ^{\ast }\chi.                             \nonumber
\end{eqnarray}

In Sect.~\ref{sec:III}, we shall need the form of the symmetric energy-momentum
tensor $T_{\mu \nu}$ of the model
\begin{eqnarray}
T_{\mu \nu }&=& 2\partial \mathcal{L}/\partial g^{\mu \nu }-g_{\mu \nu}
\mathcal{L}                                                           \nonumber
\\
&& = -F_{\mu \lambda }F_{\nu }^{\;\lambda }+\frac{1}{4}g_{\mu
\nu }F_{\lambda \rho }F^{\lambda \rho }                               \nonumber
\\
&&+\left( D_{\mu }\phi \right) ^{\ast }D_{\nu }\phi +\left( D_{\nu }\phi
\right) ^{\ast }D_{\mu }\phi                                          \nonumber
\\
&&-g_{\mu \nu }\left( \left( D_{\mu }\phi \right) ^{\ast }D^{\mu }\phi
-V\left( \left\vert \phi \right\vert \right) \right)                  \nonumber
\\
&&+\left( D_{\mu }\chi \right) ^{\ast }D_{\nu }\chi +\left( D_{\nu }\chi
\right)^{\ast }D_{\mu }\chi                                          \nonumber
\\
&&-g_{\mu \nu }\left( \left( D_{\mu }\chi \right) ^{\ast }D^{\mu }\chi
-U\left( \left\vert \chi \right\vert \right) \right).              \label{II:9}
\end{eqnarray}
In particular, we shall need the expression for the energy density $\mathcal{E}
= T_{00}$ of a field configuration of the model
\begin{eqnarray}
\mathcal{E} &=&\frac{1}{2}E_{i}E_{i}+\frac{1}{2}B^{2}               \nonumber
\\
&&+\left( D_{0}\phi \right) ^{\ast }D_{0}\phi +\left( D_{i}\phi \right)
^{\ast }D_{i}\phi +V\left( \left\vert \phi \right\vert \right)      \nonumber
\\
&&+\left( D_{0}\chi \right) ^{\ast }D_{0}\chi +\left( D_{i}\chi \right)
^{\ast }D_{i}\chi +U\left( \left\vert \chi \right\vert \right),   \label{II:10}
\end{eqnarray}
where $E_{i} = F_{0i}$ and $B = F_{21}$ are the electric field strength and the
magnetic field strength, respectively.

\section{The vortex-Q-ball soliton system and some of its properties}
                                                                \label{sec:III}

When the  gauge  coupling constant $q$ is  equal  to  zero,  model (\ref{II:1})
possesses both the  ANO  vortex solution \cite{abrikosov,nielsen} formed of the
complex  scalar   field   $\phi$   and   the  gauge  field  $A_{\mu}$  and  the
two-dimensional nongauged  $Q$-ball solution formed of the complex scalar field
$\chi$.
The ANO vortex and the  two-dimensional $Q$-ball are electrically neutral, thus
they do not interact with each other.
The  situation  changes  drastically  if  the  gauge  coupling  constant $q$ is
different from zero.
It was shown in Ref.~\cite{loginov_plb_777}  that  in  this  case,  the soliton
system consists of interacting vortex and $Q$-ball components.
This two-dimensional vortex-Q-ball system  is  electrically neutral as a whole,
because its vortex and  $Q$-ball components have  opposite  electrical charges.
Nevertheless, the vortex-Q-ball system possesses a nonzero radial electric field
in its interior.

Using  the   Hamilton   formalism    and    Lagrange's  method  of  multipliers
\cite{FLS_1976_2,FLS_1976_3}, it was shown in Ref.~\cite{loginov_plb_777}  that
there exists  a  gauge in which  only  the  complex  scalar  field  $\chi$  has
nontrivial  time dependence  $\propto \exp\left[- i \omega t\right]$, while the
complex  scalar  field  $\phi$  and  the  Abelian  gauge field $A_{\mu}$ do not
depend on time.
It was  also  shown  that  the  vortex-Q-ball  system  satisfies  the important
differential relation
\begin{equation}
\frac{dE}{dQ_{\chi }} = \omega,                                   \label{III:1}
\end{equation}
where $E = \int\mathcal{E}d^{2}x$  and $Q_{\chi} = \int j_{\chi}^{0}d^{2}x$ are
the energy  and  the  Noether charge of the vortex-Q-ball system, respectively,
and $\omega $ is  the  phase  frequency  of  the  complex  scalar field $\chi$.
Note that in Eq.~(\ref{III:1}), the phase frequency  $\omega$  is  treated as a
function of the Noether charge $Q_{\chi}$.
Eq.~(\ref{III:1}) is a consequence of the fact that the vortex-Q-ball system is
an extremum of the energy functional $E  =  \int\mathcal{E}d^{2}x$  at  a fixed
value of the Noether charge $Q_{\chi} = \int j_{\chi}^{0}d^{2}x$.

To describe  the  vortex-Q-ball  system,  we  shall  use  the  following ansatz
\begin{subequations}                                              \label{III:2}
\begin{eqnarray}
\phi \left( r,\theta \right) & = & v\exp \left( i N \theta \right) F\left(
r\right),                                                        \label{III:2a}
\\
\chi \left( r,\theta ,t\right) & = &\exp \left[- i\left( \omega t - K \theta
\right) \right] \sigma \left( r\right),                          \label{III:2b}
\\
A^{\mu}\left( r,\theta \right) & = & \left(\frac{a_{0}\left(r\right)}{er},
\frac{1}{er}\epsilon_{ij}n_{j}a\left( r\right) \right),          \label{III:2c}
\end{eqnarray}
\end{subequations}
where $K$  and  $N$  are  integers,  $\epsilon^{ij}$  are the components of the
two-dimensional antisymmetric tensor ($\epsilon^{12}= 1$) and $n_{j}$ are those
of  the  two-dimensional  radial unit vector $\mathbf{n} = (\cos(\theta), \sin(
\theta))$.
The ansatz  functions  $a_{0}\left( r \right)$, $a\left( r \right)$, $F\left( r
\right)$,  and  $\sigma \left( r \right)$   satisfy  the  system  of  nonlinear
differential equations
\begin{eqnarray}
&&a_{0}^{\prime \prime }(r)-\frac{a_{0}^{\prime }(r)}{r}+\frac{a_{0}(r)}{
r^{2}}-2e^{2}v^{2}F\left( r\right) ^{2}a_{0}(r)                     \nonumber
\\
&&+2eqr\left[ \omega -\frac{q}{e}\frac{a_{0}(r)}{r}\right] \sigma \left(
r\right)^{2} = 0,                                                 \label{III:3}
\end{eqnarray}
\begin{eqnarray}
&&a^{\prime \prime }(r)-\frac{a^{\prime }(r)}{r}-2e^{2}v^{2}\left(
N+a(r)\right) F\left( r\right) ^{2}                                 \nonumber
\\
&&-2eq\left[ K+\frac{q}{e}a\left( r\right) \right] \sigma \left( r\right)^{2}
= 0,                                                              \label{III:4}
\end{eqnarray}
\begin{eqnarray}
&&F^{\prime \prime }(r)+\frac{F^{\prime }(r)}{r}-\frac{\left(
(N+a(r))^{2}-a_{0}(r)^{2}\right) }{r^{2}}F(r)                       \nonumber
\\
&&+\lambda v^{2}\left( 1-F(r)^{2}\right) F(r) = 0,                \label{III:5}
\end{eqnarray}
\begin{eqnarray}
&&\sigma ^{\prime \prime }(r)+\frac{\sigma ^{\prime }(r)}{r}+\left[ \left(
\omega -\frac{q}{e}\frac{a_{0}\left( r\right) }{r}\right)^{2}\right.  \nonumber
\\
&&-\left. \left( \frac{K}{r}+\frac{q}{e}\frac{a\left( r\right) }{r}\right)
^{2}\right] \sigma \left( r\right)                                    \nonumber
\\
&& - \left(m^{2}-g\sigma \left( r\right) ^{2}+h\sigma \left( r\right)
^{4}\right) \sigma \left( r\right)  = 0.                          \label{III:6}
\end{eqnarray}
The energy density of the vortex-Q-ball  system  can also be expressed in terms
of the ansatz functions
\begin{eqnarray}
\mathcal{E} &\mathcal{=}&\frac{1}{2}\frac{a^{\prime }{}^{2}}{e^{2}r^{2}}+
\frac{1}{2}\left[ \left(\frac{a_{0}}{er}
\right)^{\prime}\right]^{2}                                         \nonumber
\\
&&+v^{2}F^{\prime }{}^{2}+\frac{\left( (N+a)^{2}+a_{0}{}^{2}\right) }{r^{2}}
v^{2}F^{2}                                                          \nonumber
\\
&&+\frac{\lambda }{2}v^{4}\left( F^{2}-1\right) ^{2}+\sigma ^{\prime }{}^{2}
                                                                    \nonumber
\\
&&+\left( \omega -\frac{q}{e}\frac{a_{0}}{r}\right) ^{2}\sigma ^{2}+\left(
\frac{K}{r}+\frac{q}{e}\frac{a}{r}\right)^{2}\sigma^{2}              \nonumber
\\
&& +m^{2}\sigma^{2}-\frac{g}{2}\sigma^{4}+\frac{h}{3}\sigma^{6}.  \label{III:7}
\end{eqnarray}
The regularity  of  the  vortex-Q-ball  solution  at $r = 0$ and the finiteness
of the energy $E=2 \pi\int\nolimits_{0}^{\infty}\mathcal{E}\left(r\right) r dr$
lead to the boundary conditions for the ansatz functions
\begin{subequations} \label{III:8}
\begin{align}
&\; a_{0}(0)=0,\;\underset{r\rightarrow \infty}{\lim}a_{0}(r)=0, \label{III:8a}
\\
& \; a(0) = 0,\;\underset{r\rightarrow \infty }{\lim }a(r) = -N, \label{III:8b}
\\
& \; F(0) = 0,\;\underset{r\rightarrow \infty }{\lim }F(r) = 1,  \label{III:8c}
\\
& \; \delta_{K 0}\sigma^{\prime }(0) + \left(1 - \delta _{K 0}\right)
\sigma (0) = 0, \; \underset{r\rightarrow \infty }{\lim }\sigma (r) = 0,
                                                                 \label{III:8d}
\end{align}
\end{subequations}
where $\delta_{K 0}$ is the Kronecker symbol.

Substituting   the   power    expansions    for   the   ansatz   functions   in
Eqs.~(\ref{III:3})--(\ref{III:6}) and  taking  into account boundary conditions
(\ref{III:8}), we obtain the power expansion at the origin of the vortex-Q-ball
solution with $N = 1$.
The power expansion  of  the  ansatz  function  $a_{0}\left(r \right)$  has the
form
\begin{equation}
a_{0}\left( r\right) = a_{1}r + \frac{a_{p}}{p!}r^{p} + O\left( r^{p+2}\right),
                                                                  \label{III:9}
\end{equation}
where
\begin{equation}
p = 3\delta_{K0} + 5\left(1 - \delta _{K 0}\right)               \label{III:10}
\end{equation}
and
\begin{subequations}                                             \label{III:11}
\begin{eqnarray}
a_{3} & = &3qd_{0}^{2}\left( a_{1}q - e\omega \right),          \label{III:11a}
\\
a_{5} &=&15e^{2}v^{2}a_{1}c_{1}^{2} + 15\delta_{\left\vert K\right\vert
\,1}\left( d_{1}^{2}q\left( a_{1}q-e\omega \right) \right).     \label{III:11b}
\end{eqnarray}
\end{subequations}
The power expansions  of  the  ansatz  functions $a\left(r\right)$ and $F\left(
r \right)$ are written as
\begin{eqnarray}
a\left( r\right)  &=&\frac{b_{2}}{2!}r^{2}+\frac{b_{4}}{4!}r^{4}+O\left(
r^{6}\right),                                                   \label{III:12a}
\\
F\left( r\right)  &=&c_{1}r+\frac{c_{3}}{3!}r^{3}+O\left( r^{5}\right),
                                                                \label{III:12b}
\end{eqnarray}
where the next-to-leading order coefficients are
\begin{eqnarray}
b_{4} & = & 6 e^{2}v^{2}c_{1}^{2} + 3\delta_{K 0}q^{2}b_{2}d_{0}^{2} +
6\text{sgn}\left( K\right) \delta_{\left\vert
K\right\vert \,1}eqd_{1}^{2},                                   \label{III:13a}
\\
c_{3} & = & -\frac{3}{4}c_{1}\left( a_{1}^{2}-b_{2}+\lambda v^{2}\right).
                                                                \label{III:13b}
\end{eqnarray}
Finally, the power expansion of the ansatz function $\sigma \left( r\right)$ is
\begin{equation}
\sigma \left( r\right) =\frac{d_{\left\vert K\right\vert }}{\left\vert
K\right\vert !}r^{\left\vert K\right\vert }+\frac{d_{\left\vert K\right\vert
+2}}{\left( \left\vert K\right\vert +2\right) !}r^{\left\vert K\right\vert
+2}+O\left( r^{\left\vert K\right\vert +4}\right),               \label{III:14}
\end{equation}
where the next-to-leading order coefficient is
\begin{eqnarray}
d_{\left\vert K\right\vert + 2} &=&\frac{\left\vert K\right\vert + 2}{4e^{2}}
\left[ \left( e\left( m+\omega \right)
-qa_{1}\right) \right.                                             \nonumber
\\
&&\times \left. \left( e\left( m-\omega \right) +qa_{1}\right) + \tau_{K}
\right] d_{\left\vert K\right\vert}                              \label{III:15}
\end{eqnarray}
and the term $\tau_{K}$ is 
\begin{equation}
\tau _{K}=-e^{2}d_{0}^{2}\left(2 g - 3 h d_{0}^{2}\right) \delta
_{K0}+e q K b_{2}\left( 1 - \delta _{K 0}\right).                \label{III:16}
\end{equation}

To obtain the asymptotic form of the vortex-Q-ball  solution  as $r \rightarrow
\infty$,  we  linearize Eqs.~(\ref{III:3})--(\ref{III:6})   and   use  boundary
conditions (\ref{III:8}).
As a result, we obtain the expressions
\begin{eqnarray}
a_{0}\left( r\right)  &\sim &a_{\infty }\sqrt{m_{A}r}\exp \left(
-m_{A}r\right)  \nonumber \\
&&\times \left(1 - \frac{1}{8 m_{A} r}+O\left[ \left( \frac{1}{m_{A}r}\right)
^{2}\right] \right),                                            \label{III:17a}
\\
a\left( r\right)  &\sim &N+b_{\infty }\sqrt{m_{A}r}\exp \left(
-m_{A}r\right)   \nonumber \\
&&\times \left(1 + \frac{3}{8 m_{A} r}+O\left[ \left( \frac{1}{m_{A}r}\right)
^{2}\right] \right),                                            \label{III:17b}
\\
F\left( r\right)  &\sim &1+c_{\infty }\frac{\exp \left( -m_{\phi }r\right) }
{\sqrt{m_{\phi }r}}   \nonumber \\
&&\times \left( 1-\frac{1}{8 m_{\phi }r}+O\left[ \left( \frac{1}{m_{\phi }r}
\right) ^{2}\right] \right),                                    \label{III:17c}
\\
\sigma \left( r\right)  &\sim &d_{\infty }\frac{\exp \left( -\Delta _{\omega
}r\right) }{\sqrt{\Delta _{\omega }r}}\left( 1+\frac{4\left(K - q N/e\right)
^{2}-1}{8\Delta _{\omega }r}\right.   \nonumber \\
&&\left. +O\left[ \left( \frac{1}{\Delta _{\omega }r}\right)^{2}\right]
\right),                                                        \label{III:17d}
\end{eqnarray}
where $m_{A} = \sqrt{2} e v$ and $m_{\phi} = \sqrt{2 \lambda }v$ are the masses
of the gauge boson  and the scalar $\phi$-particle, respectively, and $\Delta_{
\omega} = \sqrt{m^{2} - \omega^{2}}$  is  the  mass  parameter that defines the
asymptotic behaviour of the scalar field $\chi$.
Note that the asymptotic  forms (\ref{III:17a})--(\ref{III:17d}) are valid only
if the  mass  of  any  of  the  three  particles (gauge boson, $\phi$-particle,
and $\chi$-particle) does not exceed the sum of the mass  of the two  remaining
particles, with  the  mass  parameter $\Delta_{\omega}$ playing the role of the
$\chi$-particle's mass.
Only  if  this  condition  is  met,  can  Eqs.~(\ref{III:3})--(\ref{III:6})  be
linearised.

From Eqs.~(\ref{III:9})--(\ref{III:16}), it follows  that  the behaviour of the
vortex-Q-ball solution in the neighbourhood of  $r = 0$  is  determined by four
independent parameters: $a_{1}$, $b_{2}$, $c_{1}$, and $d_{K}$.
At  the  same  time,  Eqs.~(\ref{III:17a})--(\ref{III:17d})  tell  us  that the
behaviour of the vortex-Q-ball solution  at spatial infinity is also determined
by  four  independent  parameters:  $a_{\infty}$,  $b_{\infty}$,  $c_{\infty}$,
and $d_{\infty}$.
The coincidence of  the  numbers of parameters that define the behaviour of the
solution at  the  origin and  at  infinity  makes   the existence of a solution
of  the   boundary   value  problem  in  Eqs.~(\ref{III:3})--(\ref{III:6})  and
(\ref{III:8}) possible.

Having  boundary  conditions  (\ref{III:8}),  we   can  obtain  the  constraint
on the Noether charges  $Q_{\phi}$  and $Q_{\chi}$ of the vortex-Q-ball system.
To  do  this,  we  rewrite  Eq.~(\ref{III:3})  (Gauss's law)  in  compact  form
\begin{equation}
\left[ r\left(\frac{a_{0}(r)}{e r}\right)^{\prime}\right]^{\prime} =
- r j_{0}\left(r\right),                                          \label{III:18}
\end{equation}
where $j_{0}\left(r\right)$ is the electric  charge  density expressed in terms
of the ansatz functions
\begin{eqnarray}
j_{0} &=& 2 q \omega \sigma \left( r \right)^{2}-\frac{2a_{0}\left( r \right)}
{e r} \left( q^{2}\sigma \left( r\right) ^{2} + e^{2}v^{2}F\left( r \right)^{2}
\right)\!.                                                       \label{III:19}
\end{eqnarray}
Then, we integrate both sides of  Eq.~(\ref{III:18}) with respect to  $r$  from
zero to infinity.
Using boundary    conditions    (\ref{III:8})    and    asymptotic  expressions
(\ref{III:9})--(\ref{III:16})   and   (\ref{III:17a})--(\ref{III:17d})  of  the
vortex-Q-ball solution, it is easily shown that the  integral  of the left-hand
side of Eq.~(\ref{III:18}) vanishes.
At the same time, the integral  of the right-hand side of Eq.~(\ref{III:18}) is
proportional to the electric charge of the vortex-Q-ball system.
It follows  that  the  electric  charge  of  the vortex-Q-ball system vanishes.
Combining this  fact  and  Eq.~(\ref{II:8}),  we  obtain  the constraint on the
Noether   charges  $Q_{\phi}$   and  $Q_{\chi}$  of  the  vortex-Q-ball  system
\begin{equation}
Q = e Q_{\phi} + q Q_{\chi} = 0,                                 \label{III:20}
\end{equation}
where the Noether  charges  are  expressed  in  terms  of  the ansatz functions
\begin{subequations}                                             \label{III:02}
\begin{eqnarray}
Q_{\phi } &=&-4\pi v^{2}\int\limits_{0}^{\infty }a_{0}\left(r \right)
F\left( r\right)^{2}dr,                                         \label{III:02a}
\\
Q_{\chi } &=&4\pi \int\limits_{0}^{\infty}\left( \omega -\frac{q}{e}
\frac{a_{0}\left(r\right)}{r}\right)\sigma\left(r\right)^{2} r dr.
                                                                \label{III:02b}
\end{eqnarray}
\end{subequations}

Gauss's  law  (\ref{III:3})  allows  us  to  ascertain  some  global properties
of the  time  component  of  the  gauge  potential $A_{0}(r) = a_{0}(r)/(e r)$.
To do this, we rewrite Eq.~(\ref{III:3}) in the form
\begin{equation}
\left( r \Omega ^{\prime }\left( r \right) \right) ^{\prime } = 2 r \left[
q^{2}\sigma \left( r\right) ^{2}\Omega \left( r\right) - e^{2}v^{2}F\left(
r\right)^{2}\left(\omega -\Omega \left(r\right) \right) \right], \label{III:21}
\end{equation}
where the function
\begin{equation}
\Omega \left( r\right) = \omega - q A_{0}(r) = \omega - \frac{q}{e}
                         \frac{a_{0}\left( r\right) }{r}.        \label{III:22}
\end{equation}
According to Eq.~(\ref{III:8a}), the function $\Omega \left(r\right)\rightarrow
\omega$ as $r \rightarrow \infty$.
Let the phase frequency  $\omega$  be positive, then from Eq.~(\ref{III:21}) it
follows that $0<\Omega\left(r\right) < \omega$.
Indeed, if $\Omega \left(r\right) < 0$   $\left( \Omega \left(r\right) > \omega
\right)$ at some $r = \bar{r}$, then $\Omega' \left(r\right)$  will be negative
(positive)   at  $r   >   \bar{r}$,  and  thus  the  boundary condition $\Omega
\left( r\right) \underset{r \rightarrow \infty }{\rightarrow }\omega$ cannot be
satisfied.
If the phase frequency $\omega$ is negative, then $\omega <\Omega\left(r\right)
< 0$.
Thus, we have the global conditions on $\Omega\left(r\right)$
\begin{eqnarray}
0 &<&\Omega \left( r\right) <\omega \;\text{if}\;\omega >0,         \nonumber
\\
\omega  &<&\Omega \left( r\right) <0\;\text{if}\;\omega <0,      \label{III:23}
\end{eqnarray}
which can be rewritten in terms  of  the  gauge  potential $A_{0}(r) = a_{0}(r)
/(e r)$ as
\begin{eqnarray}
0 & < & A_{0}\left( r\right) < \frac{\omega }{q}  \;\text{if}\;\omega >0,
                                                                    \nonumber
\\
\frac{\omega }{q} & < & A_{0}\left( r\right) < 0\;\text{if}\;\omega <0.
                                                                 \label{III:24}
\end{eqnarray}

Eq.~(\ref{III:8b}) tells us that the boundary conditions for the ansatz function
$a\left(r\right)$ are the same as those for the ANO vortex.
It follows that the magnetic flux  of  the vortex-Q-ball system is quantized as
it is for the ANO vortex
\begin{equation}
\Phi = 2 \pi \int\limits_{0}^{\infty} B\left( r \right) rdr = \frac{2 \pi}{e}N,
                                                                 \label{III:25}
\end{equation}
where $B(r)= - a^{\prime }(r)/(e r)$ is the magnetic field strength.
In particular, the magnetic flux of the vortex-Q-ball system does not depend on
the gauge coupling constant $q$  that  determines  the  strength of interaction
between the gauge field $A_{\mu}$ and the complex scalar field $\chi$.

Having the symmetric  energy-momentum  tensor  (\ref{II:9}),  we  can  form the
angular momentum tensor
\begin{equation}
J^{\lambda \mu \nu }=x^{\mu }T^{\lambda \nu }-x^{\nu }T^{\lambda \mu }.
                                                                 \label{III:26}
\end{equation}
Use of  Eqs.~(\ref{II:9}),  (\ref{III:2}),  and  (\ref{III:26})  results in the
angular momentum density $\mathcal{J}=\frac{1}{2}\epsilon _{ij}J^{0ij}=J^{012}$
expressed in terms of the ansatz functions
\begin{eqnarray}
\mathcal{J} &=&-rBE_{r}+2\left( K+\frac{q}{e}a\right) \left( \omega -\frac{q
}{e}\frac{a_{0}}{r}\right) \sigma ^{2}                                \nonumber
\\
&&-2\frac{a_{0}\left(N + a\right)}{r}v^{2}F^{2},                 \label{III:27}
\end{eqnarray}
where $E_{r}(r)=-\left(a_{0}\left( r\right) /\left( er\right) \right)^{\prime}$
is the radial  electric field strength.
Next, we integrate the term $-r B E_{r}=-e^{-2}a^{\prime}\left(a_{0}/r\right)^{
\prime}$ by parts,  taking  into  account boundary conditions (\ref{III:8}) and
using Gauss's law (\ref{III:3}) to eliminate $a_{0}^{\prime \prime}$.
As a  result,  the   expression   for   the   angular   momentum  $J  =  2  \pi
\int\nolimits_{0}^{\infty }\mathcal{J}\left( r\right) rdr$ of the vortex-Q-ball
system takes the form
\begin{eqnarray}
J &=&-4\pi N\int\limits_{0}^{\infty }v^{2}a_{0}\left( r\right) F\left(
r\right) ^{2}dr                                                       \nonumber
\\
&&+4\pi K\int\limits_{0}^{\infty }r\left( \omega -\frac{q}{e}\frac{
a_{0}\left( r\right) }{r}\right) \sigma \left( r\right)^{2}dr         \nonumber
\\
&=&N Q_{\phi} + K Q_{\chi},                                      \label{III:28}
\end{eqnarray}
where the last line  in  Eq.~(\ref{III:28})  follows  from Eqs.~(\ref{III:02a})
and (\ref{III:02b}).
Using Eq.~(\ref{III:20}), we can rewrite Eq.~(\ref{III:28})  in  two equivalent
forms
\begin{equation}
J=\left(K  - \frac{q}{e}N\right)Q_{\chi}=\left(N - \frac{e}{q}K\right)Q_{\phi}.
                                                                 \label{III:29}
\end{equation}

Let suppose that the gauge coupling constants $e$ and $q$ are multiples of some
minimal gauge coupling  constant,  then  the  ratio $q/e$ is a rational number.
Eq.~(\ref{III:29}) tells  us  that  in  this  case, the angular momentum of the
vortex-Q-ball system vanishes when the ratio $K/N$ is equal to $q/e$.
Conversely,  the  ratio  $q/e$   is   a  rational  number  if  there  exists  a
vortex-Q-ball system possessing zero angular momentum.
Note that if $K/N = q/e$  then  field  configuration (\ref{III:2}) is invariant
under an axial rotation modulo  the  corresponding  gauge transformation, as it
should be for a gauged soliton system possessing zero angular momentum.

Using  Eqs.~(\ref{III:3})--(\ref{III:7}),  (\ref{III:02}),  and (\ref{III:28}),
one can easily  ascertain  properties  of  the  vortex-Q-ball  system under the
change of sign of the phase frequency
\begin{subequations}                                             \label{III:30}
\begin{eqnarray}
E\left( \omega \right)  &=&E\left( -\omega \right),             \label{III:30a}
\\
Q_{\phi ,\chi }\left( \omega \right)  &=&-Q_{\phi ,\chi }\left( -\omega
\right) ,                                                       \label{III:30b}
\\
J\left( \omega \right)  &=&-J\left( -\omega \right),            \label{III:30c}
\end{eqnarray}
\end{subequations}
and under the change of signs of the winding numbers $N$ and $K$
\begin{subequations}                                             \label{III:31}
\begin{eqnarray}
E\left( N,K\right)  &=&E\left( -N,-K\right),                    \label{III:31a}
\\
Q_{\phi ,\chi }\left( N,K\right)  &=&Q_{\phi ,\chi }\left( -N,-K\right),\;
                                                                \label{III:31b}
\\
J\left( N,K\right)  &=&-J\left( -N,-K\right).                   \label{III:31c}
\end{eqnarray}
\end{subequations}

From Eq.~(\ref{III:7}) it follows that the  energy of the vortex-Q-ball  system
can  be  presented  as the sum of five terms
\begin{equation}
E=E^{\left( E\right) }+E^{\left( B\right) }+E^{(G)}+E^{(T)}+E^{\left(
P\right) },                                                      \label{III:32}
\end{equation}
where
\begin{equation}
E^{\left( E\right) }=\frac{1}{2}\int E_{i}E_{i}d^{2}x =
2\pi \int\limits_{0}^{\infty} \frac{1}{2}\left[ \left(
\frac{a_{0}}{er}\right) ^{\prime }\right]^{2}rdr                 \label{III:33}
\end{equation}
is the energy of the electric field,
\begin{equation}
E^{\left( B\right) } = \frac{1}{2}\int B^{2} d^{2}x =
2\pi \int\limits_{0}^{\infty }
\frac{1}{2}\left( \frac{a^{\prime }{}}{er}\right)^{2}rdr         \label{III:34}
\end{equation}
is the energy of the magnetic field,
\begin{equation}
E^{\left( P\right)}=\int\limits_{0}^{\infty }
\left[ V\left( \left\vert \phi \right\vert \right)
+U\left( \left\vert \chi \right\vert \right) \right] d^{2}x      \label{III:35}
\end{equation}
is the potential part of the  energy,
\begin{eqnarray}
E^{\left( G\right) } &=&2\pi \int\limits_{0}^{\infty}
\left[ v^{2}F^{\prime }{}^{2}+\frac{(N+a)^{2}}{r^{2}}v^{2}F^{2}\right.
\nonumber
\\
&&\left. +\sigma ^{\prime }{}^{2}+\left( \frac{K}{r}+\frac{q}{e}\frac{a}{r}
\right) ^{2}\sigma ^{2}\right] rdr                               \label{III:36}
\end{eqnarray}
is the gradient part of the  energy, and
\begin{equation}
E^{\left( T\right) } = 2 \pi \int\limits_{0}^{\infty }
\left[ \left( \omega -\frac{q}{e}\frac{a_{0}
}{r}\right)^{2}\sigma^{2}+\frac{a_{0}{}^{2}}{r^{2}}v^{2}F^{2}\right] rdr
                                                                 \label{III:37}
\end{equation}
is the kinetic part of the  energy.
In Ref.~\cite{loginov_plb_777}, it  was  shown  that the parts of the energy of
the vortex-Q-ball system satisfy the virial relation
\begin{equation}
2\left( E^{\left( E\right) }-E^{\left( B\right) }+E^{\left( P \right)
}\right) - \omega Q_{\chi } = 0.                                 \label{III:38}
\end{equation}

The energy of the  vortex-Q-ball system  can  be  written in several equivalent
forms.
For this, we  integrate  the energy density of the electric field $\left[\left(
A_{0}/(er)\right)^{\prime}\right]^{2}$ by parts using Gauss's law (\ref{III:3})
and taking into account the boundary conditions (\ref{III:8a}).
As  a  result,  we  obtain  the  following  expression  for  the  energy of the
vortex-Q-ball system
\begin{equation}
E=\frac{\omega }{2}Q_{\chi}+E^{\left(B\right) }+E^{(G)}+E^{(P)}. \label{III:39}
\end{equation}
Combining Eqs.~(\ref{III:32}) and (\ref{III:39}), we  obtain the expression for
the Noether  charge $Q_{\chi}$ in terms of $E^{\left(E\right) }$ and $E^{\left(
T\right)}$
\begin{equation}
 Q_{\chi } = 2 \omega^{-1} \left(E^{( E )} + E^{(T)} \right).    \label{III:40}
\end{equation}
Next, Eqs.~(\ref{III:38}) and (\ref{III:40})  result  in  the  linear  relation
between  the  parts  of the energy of the vortex-Q-ball system
\begin{equation}
E^{(T)}+E^{\left( B\right) }-E^{(P)} = 0.                        \label{III:41}
\end{equation}

Eqs.~(\ref{III:38}) and (\ref{III:40}) form  a  system of two linear equations.
Using this system,  we can express the pairs of variables $(E^{(E)}, E^{(B)})$,
$(E^{(E)},  E^{(P)})$,  $(E^{(T)}, E^{(B)})$, and $(E^{(T)}, E^{(P)})$ in terms
of corresponding remaining variables.
Substituting  these  expressions  in  Eq.~(\ref{III:32}),  we  obtain four more
expressions for the energy of the vortex-Q-ball system
\begin{subequations}                                             \label{III:42}
\begin{eqnarray}
E &=&\frac{\omega }{2}Q_{\chi }+E^{(G)}+2E^{(P)}-E^{\left( T\right)}
                                                                \label{III:42a}
\\
&=&\frac{\omega }{2}Q_{\chi }+2E^{\left( B\right) }+E^{(G)}+E^{(T)}
                                                                \label{III:42b}
\\
&=&E^{\left( E\right) }+E^{(G)}+2E^{(P)}                        \label{III:42c}
\\
&=&\omega Q_{\chi }-E^{\left( E\right) }+E^{(G)}+2E^{(B)}.      \label{III:42d}
\end{eqnarray}
\end{subequations}

The spatial components  of the energy-momentum tensor are expressed in terms of
the ansatz functions as follows:
\begin{equation}
T_{ij}=\left( \frac{x_{i}x_{j}}{r^{2}}-\frac{1}{2}\delta _{ij}\right)
s\left( r\right) +\delta _{ij} p\left( r\right),                 \label{III:43}
\end{equation}
where the radial functions
\begin{eqnarray}
s\left( r\right) &=&-\left[\left(\frac{a_{0}}{er}\right)^{\prime}\right]^{2}
-2\frac{(N+a)^{2}}{r^{2}}v^{2}F^{2}                                \nonumber
\\
&&-2\left( \frac{K}{r}+\frac{q}{e}\frac{a}{r}\right) ^{2}\sigma ^{2}+2\left(
v^{2}F^{\prime }{}^{2}+\sigma ^{\prime }{}^{2}\right)            \label{III:44}
\end{eqnarray}
and
\begin{eqnarray}
p\left( r\right)  &=&\frac{1}{2}\frac{a^{\prime }{}^{2}}{e^{2}r^{2}}+\left(
\omega -\frac{q}{e}\frac{a_{0}}{r}\right) ^{2}\sigma ^{2}    \nonumber
\\
&&+\frac{a_{0}{}^{2}}{r^{2}}v^{2}F^{2}-V\left( vF\right) -U\left(\sigma
\right)                                                          \label{III:45}
\end{eqnarray}
are the distribution of shear force and pressure, respectively.
The conservation of the energy-momentum tensor $\partial_{i}T^{ik} = 0$ results
in the differential relation  between  the  shear force $s\left( r \right)$ and
the pressure $p\left( r\right)$
\begin{equation}
s^{\prime }(r) + \frac{2}{r}s(r) + 2p^{\prime }(r) = 0.          \label{III:46}
\end{equation}
To obtain the Laue condition \cite{laue,birula}  for  the pressure distribution
\begin{equation}
\int\nolimits_{0}^{\infty} r p(r) dr = 0,                        \label{III:47}
\end{equation}
we multiply Eq.~(\ref{III:46}) by $r^{2}$  and integrate by parts over $r$ from
zero to infinity.
It can be  shown  that  the  Laue  condition  (\ref{III:47})  is  equivalent to
virial relation (\ref{III:38}).

\section{Extreme regimes of the vortex-Q-ball system}            \label{sec:IV}

In this section, we  ascertain  the  properties  of the vortex-Q-ball system in
several extreme regimes.
First, we consider the behaviour of the vortex-Q-ball system  at extreme values
of gauge  coupling  constants,  then  we  discuss  the  vortex-Q-ball system at
extreme values of phase  frequency $\omega$ (thick-wall and thin-wall regimes).

The two scalar fields $\phi$ and $\chi$ of  the  vortex-Q-ball  system interact
with the Abelian gauge field $A_{\mu}$.
The intensity of this  interaction  is  determined  by  the  two gauge coupling
constants $e$ and $q$.
Note that in  the  natural  units  $\hbar = c = 1$, the electric charges of the
scalar $\phi$  and  $\chi$-particles  also  equal $e$  and  $q$,  respectively.
We  suppose  that  the  group  associated  with  the  gauge  symmetry  of model
(\ref{II:1}) is the compact Abelian group $U(1)$.
In this case, the  electric  charges (gauge coupling constants) $e$ and $q$ are
commensurable and thus the ratio $\tau = q/e$ is a rational number.
This means that some  minimal  elementary  electric  charge exists and that all
others charges are multiples of this elementary charge.

Let us consider the extreme  regime  in  which  both  $e$ and $q$ tend to zero,
while the ratio $\tau = q/e$ remains finite.
It can be ascertained both analytically  and  numerically  that  as $q = e \tau
\rightarrow 0$, the time component of gauge potential
\begin{equation}
A_{0}\left(r\right) = \frac{a_{0}\left( r\right) }{er} \rightarrow \alpha_{0}
\left( r\right) e = \alpha_{0} \left( r\right) \tau ^{-1} q,       \label{IV:1}
\end{equation}
where $\alpha_{0}\left(r\right)$  is  some  function of $r$ that remains finite
as $q = e \tau \rightarrow 0$.
It follows from  Eq.~(\ref{IV:1}) that $a_{0}(r)/(e r)$ uniformly tends to zero
in the limit $q = e \tau \rightarrow 0$.
Hence, the  energy  of  the  electric  field  (\ref{III:33}) also tends to zero
($E^{(E)} \propto e^{2}$). 
Next, after the change of radial variable $r = \rho /m_{A}=\rho /(\sqrt{2}ev)$,
Eq.~(\ref{III:4}) will not explicitly depend  on  the  infinitesimal parameters
$e$ and $q$.
Instead, it  will  depend  only  on  the  ratio  $\tau = q/e$, which is finite.
On the other hand, after the change of  radial  variable $r = \rho/m_{A} = \rho
/(\sqrt{2} e v)$, variations in the  ansatz  functions  $\tilde{F}(\rho) \equiv
F(\rho/(\sqrt{2}ev))$ and $\tilde{\sigma}(\rho)\equiv\sigma(\rho/(\sqrt{2}ev))$
will concentrate  in  a  small neighbourhood of $\rho = 0$.
Hence, in the limit $q = \tau e \rightarrow 0$,  the  ansatz functions $\tilde{
F}(\rho)$ and $\tilde{\sigma}(\rho)$  in  Eq.~(\ref{III:4})  can be replaced by
their limiting values of $1$ and $0$, respectively.
After that, Eq.~(\ref{III:4}) takes the simple form
\begin{equation}
\tilde{a}^{\prime \prime }(\rho )-\frac{\tilde{a}^{\prime }(\rho )}{\rho} -
\left(N + \tilde{a}(\rho)\right) = 0,                             \label{IV:2}
\end{equation}
where $\tilde{a}(\rho) \equiv a(\rho/(\sqrt{2}ev)) = a(r)$.
The solution of Eq.~(\ref{IV:2}) satisfying  boundary  condition (\ref{III:8b})
can be written as
\begin{eqnarray}
\tilde{a}(\rho ) &=&N\left( \rho K_{1}\left( \rho \right)-1\right)  \nonumber
\\
&=&N\left[\sqrt{2}e v r K_{1}\left( \sqrt{2}evr\right) -1\right],  \label{IV:3}
\end{eqnarray}
where $K_{1}\left( \rho \right)$ is the  modified Bessel function of the second
kind.
From Eq.~(\ref{IV:3}) it follows that in terms of the initial variable $r$, the
ansatz function $a(r)$ spreads  out  over  the  interval $r \in \left[0, \infty
\right)$ as $q = \tau e \rightarrow 0$.
Specifically, for any finite $r =\bar{r}$, $a\left( \bar{r} \right) \rightarrow
0$ in this limit.
At  the  same  time,  the  ansatz   functions  $F(r)$  and  $\sigma(r)$  remain
concentrated in a finite neighbourhood of the origin as $q = \tau e \rightarrow
0$  because  Eq.~(\ref{III:5})  does   not  explicitly  depend   on  the  gauge
coupling constants, and  Eq.~(\ref{III:6}) depends  on  them  only  through the
combination $\tau = q/e$, which is finite.
These facts and Eq.~(\ref{IV:1}) lead  us  to  conclude  that  the  gauge field
$A_{\mu}$ decouples from the  Q-ball  component  of  the  vortex-Q-ball system,
and thus the Q-ball component tends to the nongauged Q-ball solution as $q=\tau
e \rightarrow 0$.
Hence, the energy of the Q-ball component  tends  to  the  finite energy of the
two-dimensional nongauged Q-ball in the limit $q = \tau e \rightarrow 0$.

The situation is  different  for  the  vortex component, which is topologically
nontrivial.
The topological nontriviality  prevents  decoupling of the gauge field from the
complex scalar field $\phi$ as $q = \tau e \rightarrow 0$.
Indeed, it follows from Eqs.~(\ref{III:34}) and (\ref{IV:3}) that the energy of
the magnetic field remains finite and tends to $\pi v^{2}N^{2}$  as $q = \tau e
\rightarrow 0$.
Moreover, it follows from Eq.~(\ref{III:25}) that the magnetic flux diverges as
$e^{-1}$ in the limit $q = \tau e \rightarrow 0$.
Nevertheless, the scalar  field  $\phi$ of the vortex  component  tends  to the
nongauged global vortex solution \cite{neu} because  in  the  limit $q = \tau e
\rightarrow 0$, the ansatz function $a(r)$ uniformly tends to zero in the region
where the ansatz function $F(r)$ differs appreciably from the limiting value of
$1$.

According to Derrick's theorem  \cite{derrick}, the energy of the global vortex
is infinite, hence the energy  of  the  vortex  component  of the vortex-Q-ball
system tends to infinity in the limit $q = \tau e \rightarrow 0$.
Indeed, let  us  integrate  the  term $(N + a)^{2}v^{2}F^{2}r^{-2}$ in gradient
energy  (\ref{III:36}) over the region  $r \gtrsim m_{\phi}$, where  the ansatz
function $F(r)$ is close to $1$.
Using Eq.~(\ref{IV:3}) for the ansatz function $a(r)$, we obtain the expression
$2 \pi^{2} N^{2} v^{2}\ln\left( m_{\phi}/m_{A} \right) \propto \ln \left(e^{-1}
\right)$, which diverges logarithmically as $e \rightarrow 0$.

Now we consider another extreme regime in which  both  gauge coupling constants
$e$ and $q$ tend to infinity: $q=\tau e\rightarrow \infty$.
In this regime, the gauge field $A_{\mu}$ ceases to be dynamic.
Instead, it is  expressed  in  terms  of  the  complex scalar fields $\phi$ and
$\chi$ in  the whole space,  except  for the infinitesimal neighbourhood of the
origin.
Indeed, the Lagrangian density (\ref{II:1})  is  written  in  terms  of  ansatz
functions (\ref{III:2}) as
\begin{eqnarray}
\mathcal{L} & = &\frac{1}{2}\left[ \left(\frac{a_{0}}{er}\right)^{\prime}
\right]^{2}-\frac{1}{2}\frac{a^{\prime}{}^{2}}{e^{2}r^{2}}         \nonumber
\\
&&-v^{2}F^{\prime }{}^{2}+\frac{\left(a_{0}{}^{2}-(N+a)^{2}\right) }{r^{2}}
v^{2}F^{2}                                                         \nonumber
\\
&&-\frac{\lambda }{2}v^{4}\left( F^{2}-1\right) ^{2}-\sigma ^{\prime }{}^{2}
                                                                   \nonumber
\\
&&+\left( \omega -\frac{q}{e}\frac{a_{0}}{r}\right) ^{2}\sigma ^{2}-\left(
\frac{K}{r} + \frac{q}{e}\frac{a}{r}\right)^{2}\sigma^{2}           \nonumber
\\
&&-m^{2}\sigma^{2}+\frac{g}{2}\sigma^{4}-\frac{h}{3}\sigma^{6}.    \label{IV:4}
\end{eqnarray}
We see that  as  $q  =  \tau  e  \rightarrow  \infty$,  the  first two terms in
Eq.~(\ref{IV:4}) tend  to zero (provided  that the corresponding derivatives in
Eq.~(\ref{IV:4}) are finite) and thus can be neglected.
In this  case, the  field  equations  for  the  ansatz functions $a_{0}(r)$ and
$a(r)$ become purely algebraic (terms with  derivatives  can   be  neglected in
Eqs.~(\ref{III:3}) and (\ref{III:4})), so $a_{0}(r)$  and  $a(r)$ are expressed
in terms of $F(r)$, $\sigma(r)$, and the model's parameters
\begin{subequations}                                               \label{IV:5}
\begin{eqnarray}
a_{0}\left( r\right)  & \underset{e\rightarrow \infty }{%
\longrightarrow } &r\omega \tau \frac{\sigma \left(
r\right) ^{2}}{v^{2}F\left( r\right) ^{2}+\tau ^{2}\sigma \left( r\right)
^{2}},                                                            \label{IV:5a}
\\
a\left( r\right)  &\underset{e\rightarrow \infty }{%
\longrightarrow } &-\frac{Nv^{2}F\left( r\right) ^{2}+K\tau
\sigma \left( r\right) ^{2}}{v^{2}F\left( r\right) ^{2}+\tau ^{2}\sigma
\left( r\right) ^{2}}.                                            \label{IV:5b}
\end{eqnarray}
\end{subequations}
For $K =0$, the ansatz  functions $a_{0}(r)$ and $a(r)$ uniformly tend to their
limiting values (\ref{IV:5}) on  the  interval $r \in \left[0, \infty \right)$.
At  the  same  time, Eq.~(\ref{IV:5})  cannot  be  used  in  the  infinitesimal
neighbourhood of the  origin (where $r \lesssim m_{A}^{-1} \propto e^{-1}$) for
nonzero $K$.
Indeed, it follows from Eq.~(\ref{IV:5b}) that $a(0)=-N$ if $K\ne0$ ($\sigma(0)
= 0$ in this case), but  this  contradicts  boundary  condition (\ref{III:8b}).

It  was  found  that  the  behaviour   of   the  vortex-Q-ball  system  differs
substantially for zero and nonzero $K$.
From Eq.~(\ref{IV:5b}) it follows that for $K = 0$, the  ansatz function $a(r)$
varies from $0$ to the vicinity of $-N$ on the interval $\Delta r_{A} \sim e^{0
}$, which does not depend on $e$.
Hence, we obtain the sequence  of  relations: $a \sim -b_{2}r^{2}/2$, $B =- a^{
\prime}/(er)\sim b_{2}/e$,  and  $E^{\left(B\right)} = \int\left(B^{2}/2\right)
d^{2}x \propto e^{-2}$, where Eq.~(\ref{III:12a}) is used.
It follows that for zero $K$, the energy of the magnetic field tends to zero as
$q  =  \tau  e  \rightarrow  \infty$.

On the other hand, for  nonzero  $K$, the  ansatz   function $a(r)$ varies from
$0$ to the vicinity of $-N$ on the interval $\Delta r_{A}\sim m_{A}^{-1}\propto
e^{-1}$, which shrinks to the origin as $e \rightarrow \infty$.
Thus we have the chain  of  relations: $a \sim -m_{A}^{2}r^{2} = - 2 e^{2}v^{2}
r^{2}$,  $B  = -a^{\prime }/(er) \sim  4  e v^{2}$, and $E^{\left( B \right)} =
\int \left(B^{2}/2 \right) d^{2}x  \sim  16 \pi  e^{2}v^{4} \int\nolimits_{0}^{
\Delta r_{A}}rdr \propto e^{0}$.
We see that unlike the previous case, the energy of the magnetic field tends to
a finite value as $q  =  \tau  e  \rightarrow  \infty$.
This  is  because  the  magnetic  field  strength  $B$  increases  indefinitely
($\propto e$) in the infinitesimal neighbourhood ($\Delta r_{A}\propto e^{-1}$)
of the origin.
At the same time,  magnetic  flux  (\ref{III:25})  of  the vortex-Q-ball system
tends to zero as $e^{-1}$ for both $K = 0$ and $K \ne 0$.

From Eq.~(\ref{IV:5a}) it follows that for all values of $K$, the time component
$a_{0}/(er)$ of the gauge potential $A_{\mu}$ uniformly tends to zero ($\propto
e^{-1}$) in the limit of large $e$.
Furthermore, the major  part  of  variation  of the ansatz  function $a_{0}(r)$
happens on the interval that does not depend on $e$.
Thus it follows that the energy of the electric field (\ref{III:33}) also tends
to zero ($E^{(E)} \propto e^{-2}$) in the limit $q=\tau e \rightarrow  \infty$.

Substituting Eqs.~(\ref{IV:5a})  and  (\ref{IV:5b})  in  Eq.~(\ref{III:19}), we
find that the electromagnetic  current  vanishes  as  $q  =  \tau e \rightarrow
\infty$
\begin{equation}
j^{\mu }=e\left( j_{\phi }^{\mu }+\tau j_{\chi }^{\mu }\right)
\underset{e\rightarrow \infty }{\longrightarrow} 0.                \label{IV:6}
\end{equation}
For $K = 0$, Eq.~(\ref{IV:6}) is  valid for all $r \in \left[0, \infty\right)$,
while for nonzero $K$, it is  valid  only on  the interval $\left(\Delta r_{A},
\infty \right)$, where $\Delta r_{A} \sim m_{A}^{-1}\propto e^{-1}$.

Next, let us  consider  the  thick-wall  regime  of  the  vortex-Q-ball system.
In this extreme regime, the modulus of the phase frequency tends to the mass of
the  scalar $\chi$-particle: $\left\vert  \omega  \right\vert  \rightarrow  m$.
In this case, the Q-ball component of the vortex-Q-ball system spreads over the
two-dimensional space,  so  the  amplitude $\sigma$ of the complex scalar field
$\chi$ uniformly tends to zero.
The time component  $A_{0} = a_{0}/(e r)$ of the gauge potential also uniformly
tends to zero in the thick-wall regime.
From Eqs.~(\ref{III:4}) and (\ref{III:5}) it follows that  the influence of the
Q-ball component on the vortex  component  of  the  vortex-Q-ball system can be
neglected in the thick-wall regime.
Hence, the vortex component of the vortex-Q-ball system tends to the ANO vortex
solution as $\left\vert  \omega  \right\vert  \rightarrow  m$.
At the  same  time,  the  energy and the Noether charge of the Q-ball component
tend to finite values in the thick-wall regime as they do for the corresponding
nongauged two-dimensional Q-ball \cite{lee_pang,paccetti}.

To ascertain the behaviour of the vortex-Q-ball system in the thick-wall regime,
we rescale the ansatz functions $A_{0}(r)=a_{0}(r)/(e r)$, $\sigma(r)$, and the
radial variable $r$ as follows:
\begin{equation}
\sigma (r) = m^{-1/2}\Delta _{\omega }\bar{\sigma }(\bar{r}),\;
A_{0}(r) = m^{-3/2}\Delta _{\omega }^{2}\bar{A}_{0}(\bar{r}),      \label{IV:7}
\end{equation}
where $r = \Delta_{\omega}^{-1}\bar{r}$   and  $\Delta_{\omega} = \left[m^{2} -
\omega^{2}\right]^{1/2}$.
Next we introduce a new functional $F$ that is related to the energy functional
$E = \int \mathcal{E} d^{2}x$ by the Legendre transformation
\begin{equation}
F\left( \omega \right) = E\left(Q_{\chi}\right) - \omega Q_{\chi}. \label{IV:8}
\end{equation}
It can be shown that $-F\left(\omega \right)$ is equal to the Lagrangian $L = 2
\pi \int \nolimits_{0}^{\infty }\mathcal{L} r dr$  on field configurations that
satisfy Gauss's law (\ref{III:3}).
Next, using Eq.~(\ref{IV:7}), we  write  the  functional $F(\omega)$ as the sum
of the three terms
\begin{equation}
F(\omega) = F_{0} + \Delta_{\omega }^{2}F_{2} + \Delta_{\omega}^{4}F_{4}.
                                                                  \label{IV:8b}
\end{equation}
In Eq.~(\ref{IV:8b}), $F_{0}$, $F_{2}$, and $F_{4}$ are written as the integrals
of the corresponding  densities: $F_{0} = 2\pi \int \mathcal{F}_{0}rdr$, $F_{2}
=2 \pi \int \mathcal{F}_{2}\bar{r}d\bar{r}$, and $F_{4}=2\pi \int \mathcal{F}_{
4}\bar{r}d\bar{r}$, where
\begin{eqnarray}
\mathcal{F}_{0} &=&\frac{1}{2}\frac{a^{\prime }{}^{2}}{e^{2}r^{2}}
+v^{2}F^{\prime }{}^{2}                                             \nonumber
\\
&&+\frac{(N+a)^{2}}{r^{2}}v^{2}F^{2}+\frac{\lambda }{2}v^{4}\left(
F^{2}-1\right) ^{2},                                               \label{IV:9}
\end{eqnarray}
\begin{eqnarray}
\mathcal{F}_{2} &=&m^{-1}\bar{\sigma }^{\prime }{}^{2}+m^{-1}\bar{
\sigma }{}^{2}-2^{-1}m^{-2}g\bar{\sigma }{}^{4}                       \nonumber
\\
&&+2m^{-5/2}\omega q\bar{A}_{0}\bar{\sigma }{}
^{2}-m^{-3}e^{2}v^{2}\bar{A}_{0}^{2}                                  \nonumber
\\
&&+m^{-1}\bar{r}^{-2}\left(K - \tau N\right)^{2}\bar{\sigma}^{2}, \label{IV:10}
\end{eqnarray}
and
\begin{equation}
\mathcal{F}_{4}=-2^{-1}m^{-3}\bar{A}_{0}^{\prime 2}-m^{-4}q^{2}\bar{
A}_{0}^{2}\bar{\sigma }{}^{2}+3^{-1}m^{-3}h\bar{\sigma }{}^{6}.   \label{IV:11}
\end{equation}
Note  that  in  Eq.~(\ref{IV:9}),  the  prime  means  the  differentiation with
respect to the radial variable $r$, while in Eqs.~(\ref{IV:10}) and (\ref{IV:11}),
it means the  differentiation  with  respect  to  the  rescaled radial variable
$\bar{r} = \Delta_{\omega} r$.
Note also that as $\Delta_{\omega} \rightarrow 0$, we have  replaced the ansatz
functions $a\left(r\right) = a\left(\Delta_{\omega}^{-1}\bar{r} \right)  \equiv
\bar{a}\left( \bar{r}\right)$  and $F\left(r\right)=F\left(\Delta_{\omega}^{-1}
\bar{r}\right) \equiv \bar{F}\left(\bar{r}\right)$  by  their  limiting  values
$-N$ and $1$, respectively.

Eq.~(\ref{IV:10}) does not contain the  derivative  of $\bar{A}_{0}$,  and thus
$\bar{A}_{0}$ can be expressed in terms of the  ansatz  function $\bar{\sigma}$
and the model's parameters
\begin{equation}
\frac{\partial \mathcal{F}_{2}}{\partial \bar{A}_{0}} = 0 \Rightarrow
\bar{A}_{0}\left(\bar{r}\right)  = \frac{m^{1/2}\omega q}{e^{2}v^{2}}
\bar{\sigma}^{2}\left(\bar{r}\right).                            \label{IV:11b}
\end{equation}
Substituting  Eq.~(\ref{IV:11b}) into  Eq.~(\ref{IV:10}),  we  obtain  the  new
expression for $\mathcal{F}_{2}$
\begin{eqnarray}
\mathcal{F}_{2} &=&m^{-1}\bar{\sigma}^{\prime }{}^{2}+m^{-1}\bar{
\sigma }{}^{2}+m^{-1}\bar{r}^{-2}\left( K-\tau N\right)^{2}\bar{
\sigma }{}^{2}                                                   \label{IV:11c}
\\
&&-\left(2^{-1}m^{-2}g-\tau ^{2}v^{-2}\right)\bar{\sigma }{}^{4}-
\Delta_{\omega }^{2}\tau^{2}\bar{\sigma }{}^{4}\left(
m^{2}v^{2}\right)^{-1}.                                             \nonumber
\end{eqnarray}
We see that $\mathcal{F}_{2}$ is explicitly dependent  on  the  small parameter
$\Delta_{\omega}$.
Hence, Eq.~(\ref{IV:8b}) can be rewritten as
\begin{equation}
F(\omega) = F_{0} + \Delta_{\omega}^{2}\bar{F}_{2} +
\Delta_{\omega}^{4}\bar{F}_{4},                                  \label{IV:11d}
\end{equation}
where $F_{0}$  is  the  same as in Eq.~(\ref{IV:8b}), $\bar{F}_{2} = 2 \pi \int
\mathcal{\bar{F}}_{2}\bar{r}d\bar{r}$, $\bar{F}_{4} = 2 \pi \int \mathcal{\bar{
F}}_{4}\bar{r}d\bar{r}$, and the corresponding densities are
\begin{eqnarray}
\mathcal{\bar{F}}_{2} &=&m^{-1}\bar{\sigma }^{\prime }{}^{2}+\left(
m^{-1}+m^{-1}\bar{r}^{-2}\left( K-\tau N\right) ^{2}\right)
\bar{\sigma }{}^{2}                                                 \nonumber
\\
&&-\left(2^{-1}m^{-2}g-\tau ^{2}v^{-2}\right)\bar{\sigma}{}^{4}, \label{IV:11e}
\end{eqnarray}
\begin{eqnarray}
\mathcal{\bar{F}}_{4} &=&-2^{-1}m^{-3}\bar{A}_{0}^{\prime 2}-m^{-4}q^{2}
\bar{A}_{0}^{2}\bar{\sigma }{}^{2}                                  \nonumber
\\
&&-\tau ^{2}\left( m^{2}v^{2}\right)^{-1}\bar{\sigma }{}
^{4}+3^{-1}m^{-3}h\bar{\sigma }{}^{6}.                           \label{IV:11f}
\end{eqnarray}
We see that  none  of  $F_{0}$, $\bar{F}_{2}$, and  $\bar{F}_{4}$ depend on the
phase frequency $\omega$.

Using  the  known  properties  of  the  Legendre  transformation, we obtain the
Noether charge  $Q_{\chi}$  and  the  energy $E$ of the vortex-Q-ball system as
functions of the phase frequency $\omega$
\begin{equation}
Q_{\chi }\left( \omega \right) =-\frac{dF\left( \omega \right) }{d\omega}
= 2 \bar{F}_{2}\omega + 4 \bar{F}_{4}\omega \Delta_{\omega }^{2}, \label{IV:12}
\end{equation}
and
\begin{eqnarray}
E\left( \omega \right) &=&F\left( \omega \right) -
\omega \frac{dF\left( \omega \right) }{d\omega }                  \label{IV:13}
\\
&=&F_{0}+\Delta_{\omega}^{2}\left(\bar{F}_{2}+\bar{F}_{4}\Delta_{\omega}^{2}
\right)+2\omega^{2}\left(\bar{F}_{2}+2\bar{F}_{4}\Delta_{\omega}^{2}\right).
                                                                    \nonumber
\end{eqnarray}
From Eqs.~(\ref{IV:12}) and (\ref{IV:13}) it follows  that  the  energy and the
Noether charge of the vortex-Q-ball system  tend  to  the  finite  values  $E_{
\text{tk}}=F_{0}+2\omega_{\text{tk}}^{2}\bar{F}_{2}$   and  $Q_{\chi \text{tk}}
=2\omega_{\text{tk}}\bar{F}_{2}$, respectively, as $\omega \rightarrow \omega_{
\text{tk}} = \pm m$.
Moreover, Eqs.~(\ref{IV:12}) and (\ref{IV:13}) make  it  possible to obtain the
dependence of the energy on the Noether  charge, which is valid  for both signs
of $Q_{\chi}$ in the thick-wall regime
\begin{equation}
E\left( Q_{\chi }\right) =E_{\text{tk}}-m\Delta Q_{\chi }+O\left[ \Delta
Q_{\chi }^{2}\right],                                             \label{IV:14}
\end{equation}
where  $\Delta Q_{\chi} = \left\vert Q_{\chi \text{tk}}\right\vert - \left\vert
Q_{\chi }\right\vert$.

Although the influence  of  the  Q-ball  component  on  the vortex component is
negligible in the thick-wall regime, but the reverse is not true.
Indeed, by varying the functional $\bar{F}_{2} = 2 \pi \int \mathcal{\bar{F}}_{
2}\bar{r}d\bar{r}$ in $\bar{\sigma}$,  we  obtain the differential equation for
$\bar{\sigma}$, which is valid in the thick-wall regime
\begin{gather}
\bar{\sigma }^{\prime \prime }+\frac{\bar{\sigma }^{\prime }}{
\bar{r}}-\left( 1+\bar{r}^{-2}\left( K-\tau N\right) ^{2}\right)
\bar{\sigma }{}                                                     \nonumber
\\
+ \left( m^{-1}g-2m\tau ^{2}v^{-2}\right) \bar{\sigma }^{3} = 0.  \label{IV:15}
\end{gather}
We see that Eq.~(\ref{IV:15}) depends on the  parameters  $N$  and  $v$  of the
vortex component, hence,  the  influence  of the vortex component on the Q-ball
component can not be neglected in the thick-wall regime.

Eq.~(\ref{IV:11e}) tells us that the functional $\bar{F}_{2}=2\pi\int \mathcal{
\bar{F}}_{2}\bar{r}d\bar{r}$ depends on the gauge coupling constants $e$ and $q$
only through the combination $\tau = q/e$.
Hence, the limiting value $Q_{\chi \text{tk}} = 2\omega_{\text{tk}}\bar{F}_{2}$
depends on $e$ and $q$ only through the combination $\tau = q/e$.
At the same time, we see  from  Eq.~(\ref{IV:9}) that  the  functional $F_{0} =
2 \pi \int \mathcal{F}_{0} r dr$ (the energy  of  the ANO vortex for given $e$,
$\lambda$, $v$, and $N$) depends solely  on  the  gauge  coupling constant $e$.
Hence, the limiting energy $E_{\text{tk}} = F_{0} + 2 m^{2} \bar{F}_{2}$ of the
vortex-Q-ball system depends on $e$ and $q$ separately.
In particular,  it  diverges  logarithmically  ($\propto \ln(e^{-1})$) when $e$
tends to zero and $\tau$ remains fixed.

Finally,  we  consider  the  thin-wall  regime  of  the  vortex-Q-ball  system.
This extreme regime occurs when  the  absolute  value $\left\vert \omega \right
\vert$  of  the  phase  frequency tends to the minimum possible value $\omega_{
\text{tn}}$.
As a result, the energy $E = 2 \pi \int\nolimits_{0}^{\infty } \mathcal{E} rdr$
and the Noether charge $Q_{\chi} = 2\pi \int\nolimits_{0}^{\infty }j_{\chi }^{0
}rdr$ increase indefinitely as $\omega \rightarrow \omega_{\text{tn}}$.
In the thin-wall regime, the  field  configuration  of the vortex-Q-ball system
can be divided into  three  regions: the central transitional region, the basic
interior region, and the exterior transitional region.
The  spatial  size  of  the  internal  region  increases  indefinitely  as  the
vortex-Q-ball system approaches the thin-wall limit.
The  characteristic  feature  of  the  thin-wall  regime  is  that  the  ansatz
functions  $\sigma(r)$,  $\Omega(r)  =  \omega  - \tau a_{0}(r)/r$,  and $F(r)$
tends to constant values in the internal region as $\omega \rightarrow \omega_{
\text{tn}}$.
Equating the derivatives in Eqs.~(\ref{III:3}), (\ref{III:5}), and (\ref{III:6})
to zero  (Eq.~(\ref{III:3}) should  be  rewritten in terms of $\Omega(r)$ as it
is in Eq.~(\ref{III:21})), we obtain a system of three algebraic equations
\begin{eqnarray}
\Omega ^{2}-m^{2}+g\sigma ^{2}-h\sigma ^{4} &=&0,                 \label{IV:16}
\\
\tau ^{2}\sigma ^{2}\Omega -v^{2}F^{2}(\omega -\Omega ) &=&0,     \label{IV:17}
\\
(\omega -\Omega )^{2}+\tau ^{2}\lambda v^{2}\left( 1-F^{2}\right)  &=&0.
                                                                  \label{IV:18}
\end{eqnarray}
One more equation can be obtained from Eq.~(\ref{III:41}).
Indeed, in the thin-wall regime, the energy of the magnetic field tends to zero,
while  the  potential  and  kinetic  parts  of  energy  increase  indefinitely.
Hence, the  term  $E^{(B)}$  can  be  neglected  in  the  thin-wall  regime and
Eq.~(\ref{III:41}) takes the form $E^{(T)} - E^{(P)} = 0$.
This fact  and  Eqs.~(\ref{III:35})  and  (\ref{III:37})  result  in the fourth
algebraic equation
\begin{align}
&\Omega ^{2}\sigma ^{2}+\tau ^{-2}v^{2}(\omega -\Omega)^{2}F^{2}  \label{IV:19}
\\
&= 2^{-1}\lambda v^{4}\left( 1-F^{2}\right) ^{2}+m^{2}\sigma ^{2}-2^{-1}g\sigma
^{4}+3^{-1}h\sigma ^{6}.                                            \nonumber
\end{align}

Before we go any further, let us ascertain the behaviour of the ansatz function
$a(r)$ in  the  internal  region  of  the vortex-Q-ball system in the thin-wall
regime.
From Eq.~(\ref{III:4}) it follows that in this case, the behaviour of $a(r)$ is
described  by the differential equation
\begin{equation}
a^{\prime \prime}(r) - \frac{a^{\prime}(r)}{r} - \beta a(r) - \gamma = 0,
                                                                  \label{IV:20}
\end{equation}
where
\begin{eqnarray}
\beta &=&2e^{2}\left(v^{2}F_{\text{tn}}^{2}+\tau^{2}\sigma_{\text{tn}
}^{2}\right),                                                        \nonumber
\\
\gamma  &=&2e^{2}\left( Nv^{2}F_{\text{tn}}^{2}+K\tau \sigma _{\text{tn}
}^{2}\right),                                                     \label{IV:21}
\end{eqnarray}
and $F_{\text{tn}}$  and  $\sigma_{\text{tn}}$  are  the limiting values of the
corresponding ansatz functions in the thin-wall regime.
The appropriate solution of Eq.~(\ref{IV:20}) is
\begin{equation}
a(r)=C r \text{K}_{1}\left(\sqrt{\beta}r\right)-\beta^{-1}\gamma, \label{IV:22}
\end{equation}
where $K_{1}\left(\sqrt{\beta} r\right)$ is the modified Bessel function of the
second kind and $C$ is a positive constant.
We see that in the internal region of the vortex-Q-ball system, $a(r)$ tends to
the constant value
\begin{equation}
a_{\text{tn}}=-\frac{\gamma}{\beta} =
-N - \tau K\frac{\sigma_{\text{tn}}^{2}}
{v^{2}F_{\text{tn}}^{2}} + \tau^{2}N\frac{\sigma_{\text{tn}}^{2}}
{v^{2}F_{\text{tn}}^{2}}+O\left[ \tau ^{3}\right].                \label{IV:23}
\end{equation}
Note that the limiting value (\ref{IV:23})  does  not  equal the boundary value
$-N$ of $a(r)$ at spatial infinity.

It follows from Eq.~(\ref{IV:23}) that the behaviour of $a(r)$ is determined by
the thin-wall  background  values  $F_{\text{tn}}$  and $\sigma_{\text{tn}}$ of
$F(r)$ and $\sigma(r)$, respectively.
At the  same  time,  Eqs.~(\ref{III:5})  and  (\ref{III:6}) tell us that in the
thin-wall regime, the backward influence of $a(r)$ on $F(r)$ and $\sigma(r)$ can
be neglected in the internal region of the vortex-Q-ball system.
Indeed, the ansatz functions  $a_{0}(r)$  and  $\Omega(r)$ are connected by the
relation $a_{0}=\tau^{-1}r\left( \omega - \Omega \right)$, and the constancy of
$\Omega$ in the internal region  of  the  vortex-Q-ball  system  results in the
linear growth of $a_{0}$ in $r$ there.
Conversely, the ansatz function $a(r)$ is bounded in the internal region of the
vortex-Q-ball system and the ratio $a(r)/a_{0}(r)$ tends to zero there, because
the size of the internal region increases indefinitely in the thin-wall regime.

Thus, we have the system of four algebraic equations (\ref{IV:16})--(\ref{IV:19}),
which are valid in the thin-wall regime.
These  equations  allow   us   to   determine  the  limiting  thin-wall  values
$\Omega_{\text{tn}}$,  $\sigma_{\text{tn}}$,    and    $F_{\text{tn}}$   of the
corresponding ansatz functions and the limiting thin-wall value $\omega_{\text{
tn}}$ of the phase frequency.
The  solution  of   system  (\ref{IV:16})--(\ref{IV:19})  cannot   be  obtained
analytically, in general.
However, this solution can be obtained as series in the parameter $\tau = q/e$,
provided it is small enough
\begin{subequations}                                              \label{IV:24}
\begin{eqnarray}
\omega _{\text{tn}} &=&\omega _{\text{tn}0}\left( 1+\frac{3}{8}
\frac{g}{hv^{2}}\tau ^{2}+O\left[ \tau ^{4}\right] \right),      \label{IV:24a}
\\
\Omega _{\text{tn}} &=&\omega _{\text{tn}0}\left( 1-\frac{3}{8}
\frac{g}{hv^{2}}\tau ^{2}+O\left[ \tau ^{4}\right] \right),      \label{IV:24b}
\\
\sigma _{\text{tn}} &=&\sigma _{\text{tn}0}\left( 1-\frac{\omega
_{\text{tn}0}^{2}}{gv^{2}}\tau ^{2}+O\left[ \tau ^{4}\right]
\right),                                                         \label{IV:24c}
\\
F_{\text{tn}} &=&1+\frac{9}{32}\frac{g^{2}\omega _{\text{tn}0}^{2}
}{\lambda h^{2}v^{6}}\tau ^{2}+O\left[ \tau ^{4}\right],         \label{IV:24d}
\end{eqnarray}
\end{subequations}
where
\begin{equation}
\sigma _{\text{tn}0}=\frac{\sqrt{3}}{2}\sqrt{\frac{g}{h}}\quad \text{and} \quad
\omega _{\text{tn}0}=m\sqrt{1-\frac{3}{16}\frac{g^{2}}{hm^{2}}}   \label{IV:25}
\end{equation}
are the limiting thin-wall values of $\sigma(r)$ and $\omega$ for the nongauged
Q-ball.
Note that  the  system (\ref{IV:16})--(\ref{IV:19})  does  not  depend  on  the
integers  $K$  and  $N$,  hence   the  thin-wall  values  $\Omega_{\text{tn}}$,
$\sigma_{\text{tn}}$, $F_{\text{tn}}$,  and  $\omega_{\text{tn}}$ also does not
depend on $K$  and  $N$.
Furthermore, the system (\ref{IV:16})--(\ref{IV:19}), and consequently, $\Omega_{
\text{tn}}$,  $\sigma_{\text{tn}}$,  $F_{\text{tn}}$,  and $\omega_{\text{tn}}$
depends  on  the  gauge  coupling  constants  $e$  and  $q$  only  through  the
combination $\tau = q/e$.

Using Eqs.~(\ref{IV:24a})--(\ref{IV:24d}), we  can  obtain the thin-wall values
of the energy, Noether charge, and angular momentum density
\begin{subequations}                                              \label{IV:26}
\begin{eqnarray}
\mathcal{E}_{\text{tn}} &=&\mathcal{E}_{\text{tn}0}\left( 1-2
\frac{\omega _{\text{tn}0}^{2}}{gv^{2}}\tau ^{2}+O\left[ \tau ^{4}
\right] \right) ,                                                \label{IV:26a}
\\
j_{\chi \text{tn}}^{0} &=&j_{\chi \text{tn}0}^{0}\left( 1-2\frac{m^{2}}{
gv^{2}}\tau ^{2}+O\left[ \tau ^{4}\right] \right) ,              \label{IV:26b}
\\
j_{\phi \,\text{tn}}^{0} &=&-\tau j_{\chi \text{tn}0}^{0}
\left(1 + O\left[\tau^{2}\right] \right),                        \label{IV:26c}
\\
\mathcal{J}_{\text{tn}} &=&Nj_{\phi\,\text{tn}}^{0}+Kj_{\chi\,\text{tn}
}^{0},                                                           \label{IV:26d}
\end{eqnarray}
\end{subequations}
where
\begin{equation}
\mathcal{E}_{\text{tn}0}=2\omega_{\text{tn}0}^{2} \sigma_{\text{tn}0}^{2}
                          \quad \text{and} \quad
j_{\chi \text{tn}0}^{0}=2\omega _{\text{tn}0}\sigma_{\text{tn}0}^{2}
                                                                  \label{IV:27}
\end{equation}
are the corresponding densities for the nongauged Q-ball.
It follows  from  Eqs.~(\ref{IV:24a}), (\ref{IV:26a}),  and (\ref{IV:26b}) that
the ratio
\begin{equation}
\frac{\mathcal{E}_{\text{tn}}}{j_{\chi \text{tn}}^{0}}=
\omega _{\text{tn}0}\left( 1+\frac{3}{8}\frac{g}{hv^{2}}\tau ^{2} +
O\left[ \tau^{4}\right] \right) = \omega_{\text{tn}}              \label{IV:28}
\end{equation}
as it should be in the thin-wall regime.

\section{Numerical results}                                       \label{sec:V}

The  system   of   differential   equations  (\ref{III:3})--(\ref{III:6})  with
boundary conditions (\ref{III:8}) is  a  mixed  boundary  value  problem on the
semi-infinite interval $r \in \left[0, \infty \right)$.
It is obviously that this problem can  be  solved  only  by  numerical methods.
To solve this problem, we  use the boundary value  problem  solver  provided in
the {\sc{Maple}} package \cite{maple}.
The  correctness   of   numerical  results  is  controlled  with  the  help  of
Eq.~(\ref{III:1}) and the Laue condition (\ref{III:47}).

\begin{figure}[tbp]
\includegraphics[width=0.5\textwidth]{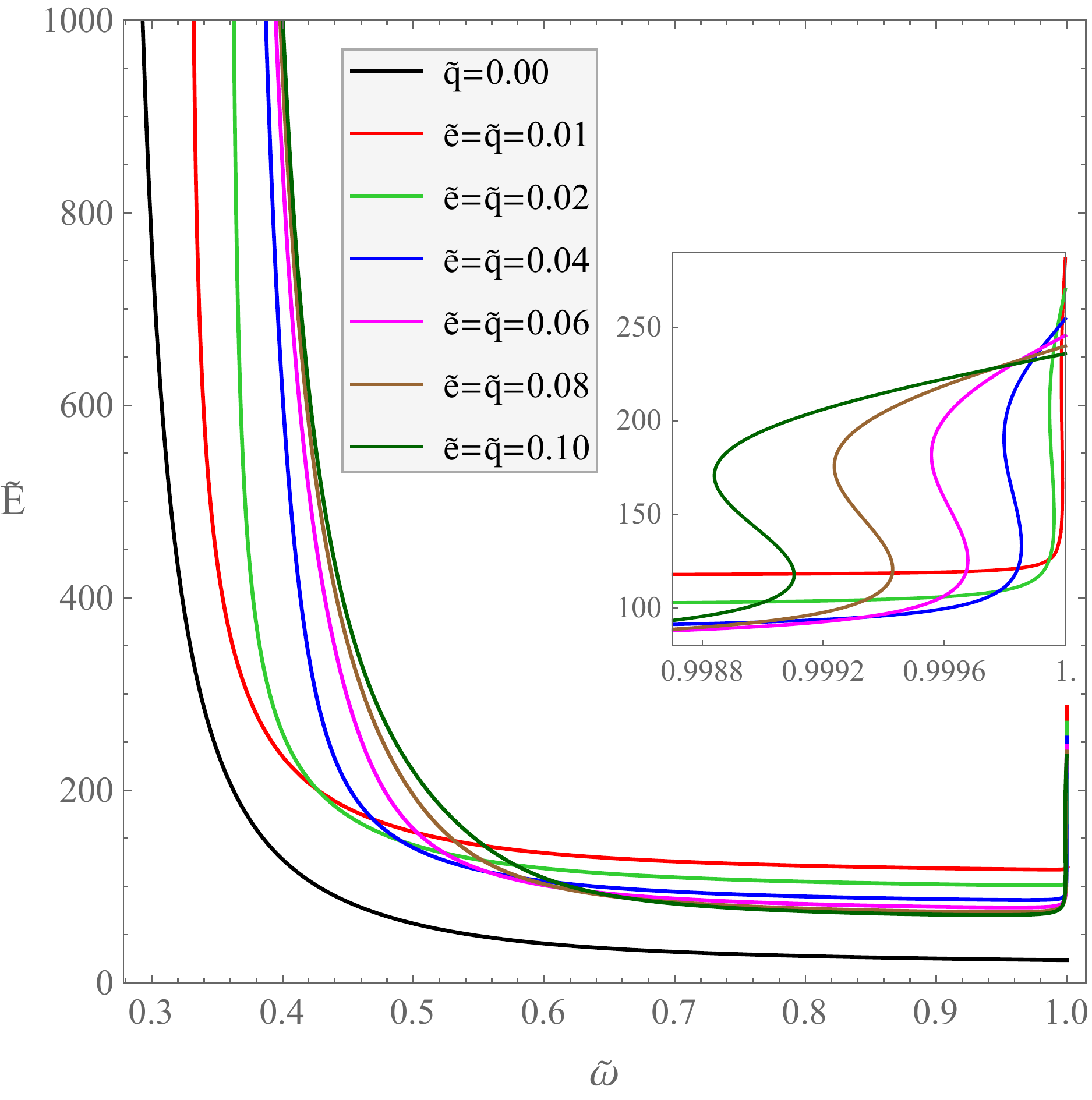}
\caption{Dependence of the energy $\tilde{E}$  of  the  vortex-Q-ball system on
the phase frequency $\tilde{\omega}$ for  different  values  of  gauge coupling
constants.  The  curves  correspond  to  parameters $\tilde{m} = 1$, $\tilde{g}
= 1$, $\tilde{h} = 0.2$,  $\tilde{\lambda} = 0.125$,  $\tilde{v} = 2$, $N = 1$,
and $K = 0$}
\label{fig:1}
\end{figure}

The mixed boundary value problem (\ref{III:3})--(\ref{III:6}), and (\ref{III:8})
depends on the  eight  parameters: $m$, $g$, $h$, $\lambda$, $v$, $e$, $q$, and
$\omega$.
Without loss of  generality, the number of the parameters can be reduced, if we
rescale the  radial  variable $r$  and the ansatz function $\sigma$ as follows:
\begin{equation}
r=m^{-1}\tilde{r},\;\sigma = mg^{-1/2}\tilde{\sigma}.               \label{V:1}
\end{equation}
After  rescaling, the  vortex-Q-ball  system  will  be  described solely by the
six  dimensionless  parameters: $\tilde{h} =  m^{2}g^{-2}h$, $\tilde{\lambda} =
g v^{4} m^{-4}\lambda$, $\tilde{v} = g^{1/2}m^{-1}v$, $\tilde{e} = g^{-1/2} e$,
$\tilde{q} = g^{-1/2}q$, and $\tilde{\omega} = m^{-1} \omega$, whereas the  two
dimensionless parameters $\tilde{m}$  and  $\tilde{g}$  will be equal to 1.
In most numerical calculations,  we  use the following values for the nongauged
dimensionless   parameters:   $\tilde{h}  =  0.2$, $\tilde{\lambda}  =  0.125$,
$\tilde{v} = 2$.
The dimensionless gauge coupling constants  $\tilde{e}$   and  $\tilde{q}$  are
taken to be equal ($\tau=\tilde{q}/\tilde{e}=1$), and may vary in some interval.
In  addition   to   these   parameters,  the   mixed   boundary  value  problem
(\ref{III:3})--(\ref{III:6}),   and  (\ref{III:8})  also  depends  on  the  two
integers: $N$  (topological  winding  number  of the vortex component)  and $K$
(nontopological winding number of the Q-ball component).
We shall consider the vortex-Q-ball systems with the vortex winding number $N =
1$, while the Q-ball winding number $K$ may take a range of values.

\begin{figure}[tbp]
\includegraphics[width=0.5\textwidth]{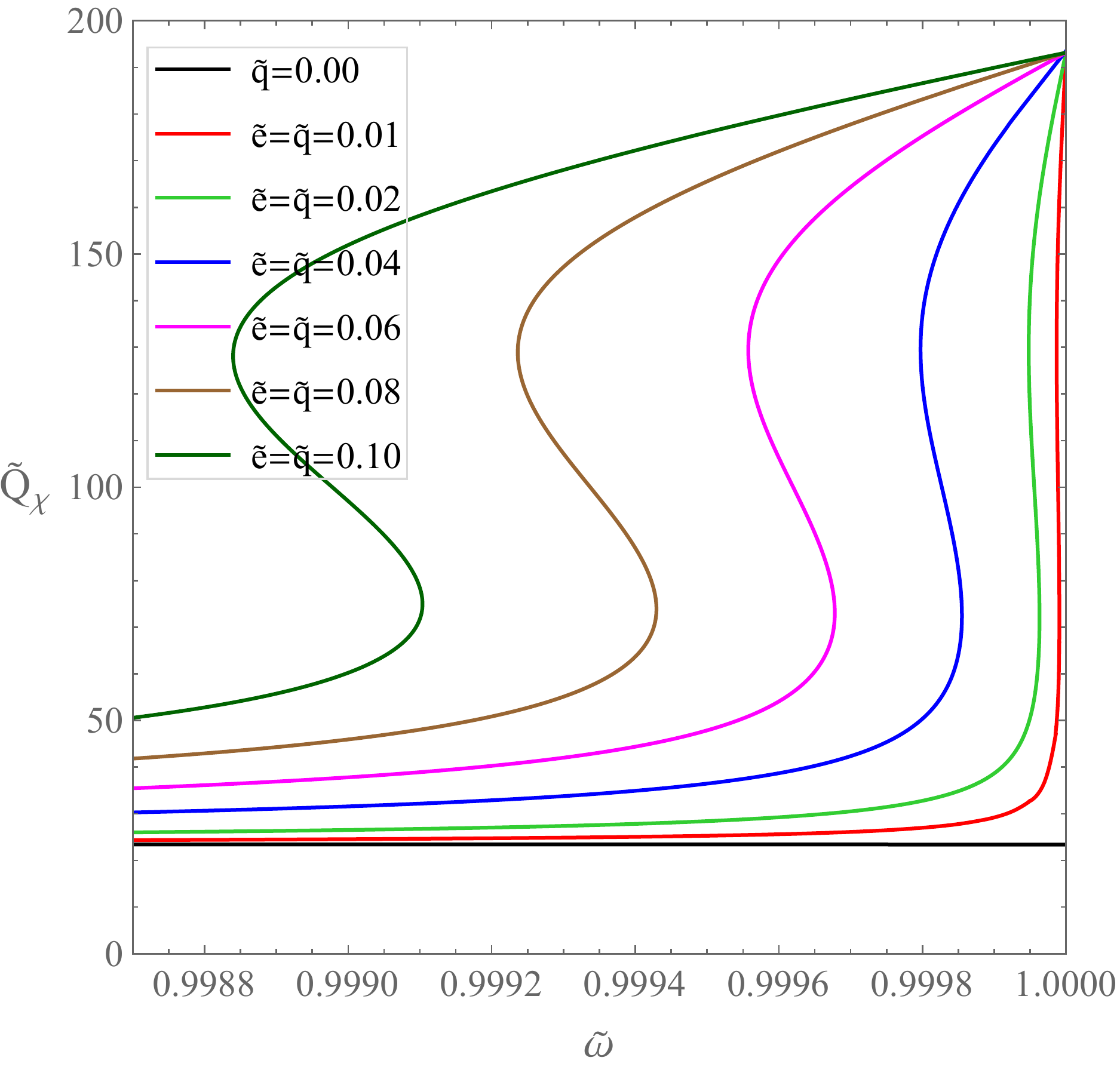}
\caption{Dependence of the Noether charge $\tilde{Q}_{\chi}$ of the vortex-Q-ball
system  on the  dimensionless  phase frequency $\tilde{\omega}$ in the vicinity
of $\tilde{\omega} = 1$.     The parameters of the vortex-Q-ball system are the
same as in Fig.~\ref{fig:1}}
\label{fig:2}
\end{figure}

Figure~\ref{fig:1} shows the dependence of the  dimensionless energy $\tilde{E}
= g m^{-2} E$  of the vortex-Q-ball system on the dimensionless phase frequency
$\tilde{\omega} = m^{-1} \omega$.
The presented curves correspond to the vortex-Q-ball  system  at  six different
values of the  gauge  coupling  constants  and to the nongauged two-dimensional
Q-ball.
We see  that  with  decreasing  $\tilde{\omega}$,  the  system  passes into the
thin-wall  regime   in  which  the  energy  and  the  Noether  charge  increase
indefinitely.
When $\tilde{\omega} \rightarrow 1$, the  vortex-Q-ball  system passes into the
thick-wall regime.
In this regime, the behaviour  of the vortex-Q-ball system differs considerably
from that of the nongauged two-dimensional Q-ball.
In  particular,  the   energy   of   the   Q-ball  decreases  monotonically  as
$\tilde{\omega} \rightarrow 1$,  whereas  the  behaviour of  the  energy of the
vortex-Q-ball system  is  more  complicated,  as   follows  from the subplot in
Fig.~\ref{fig:1}.
We see that for the vortex-Q-ball system, the curves $\tilde{E}(\tilde{\omega})$
are $s$-shaped in the vicinity of $\tilde{\omega} = 1$.
This fact is due to the  nontrivial  interaction  between the vortex and Q-ball
components of the soliton system.
Note that in  Ref.~\cite{loginov_plb_777}, we  have not  managed  to obtain, by
numerical methods, the $s$-shaped parts  of the $\tilde{E}(\tilde{\omega})$ and
$\tilde{Q}_{\chi}(\tilde{\omega})$ curves.

It follows from the subplot in Fig.~\ref{fig:1}  that the limit value $\tilde{E
}(1)$ is not a constant  for  the  vortex-Q-ball  system,  but increases with a
decrease of $\tilde{e}$.
It was found  numerically  that  for  small  values  of  $\tilde{e}$, the value
$\tilde{E}(1) \approx a + b \ln(\tilde{e}^{-1})$, where $a$ and  $b$  depend on
the gauge  coupling  constants  through  the  ratio $\tau= \tilde{q}/\tilde{e}$
in accordance with the conclusion of Sect.~\ref{sec:IV}.

\begin{figure}[tbp]
\includegraphics[width=0.5\textwidth]{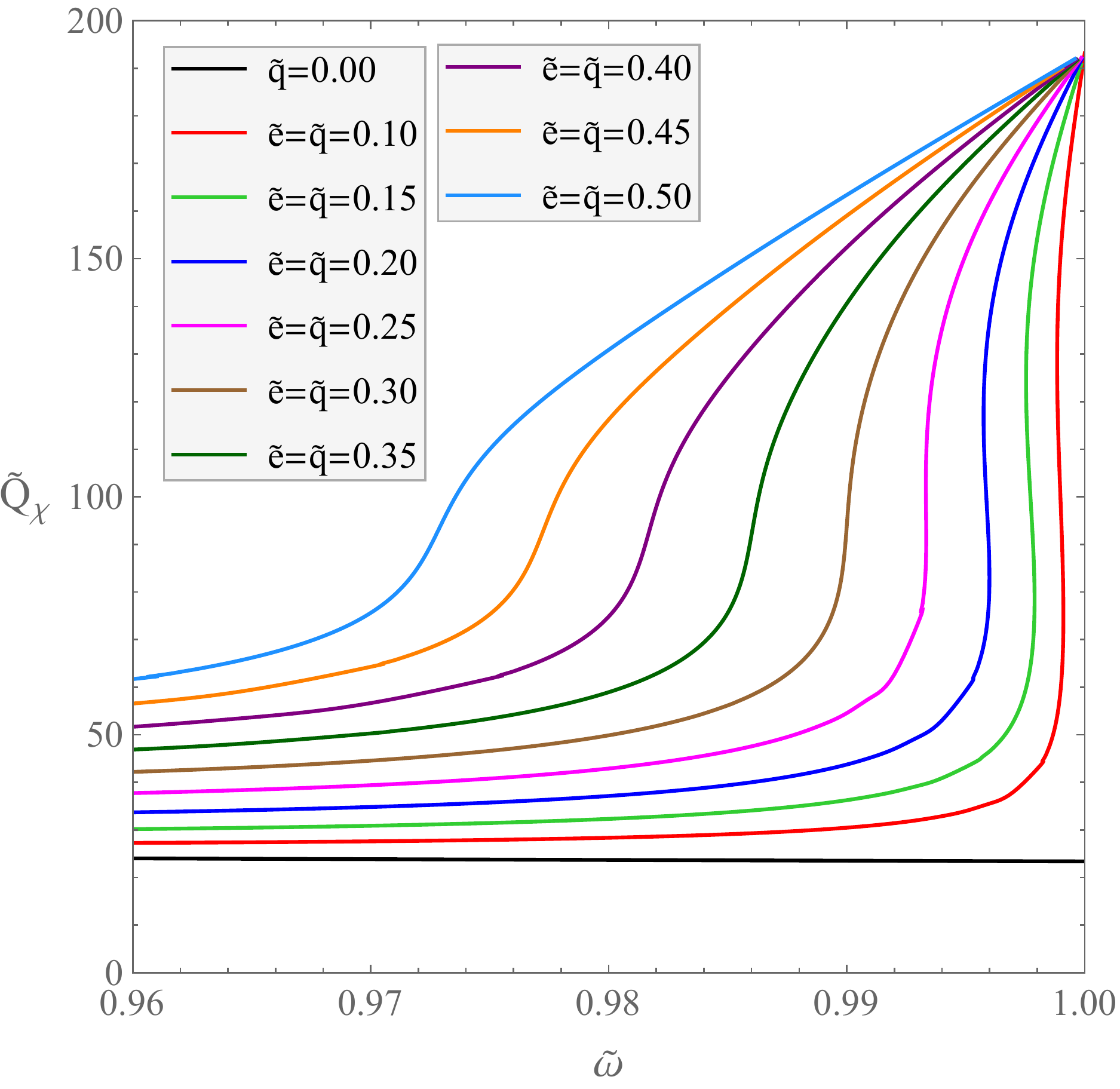}
\caption{The same as in Fig.~\ref{fig:2},  but  for  large  values of the gauge
coupling constants}
\label{fig:3}
\end{figure}

Beyond the neighbourhood  of  $\tilde{\omega} = 1$, the rescaled Noether charge
$\tilde{Q}_{\chi}=g m^{-1} Q_{\chi}$ depends on $\tilde{\omega}$ similar to the
dimensionless energy  $\tilde{E} = g m^{-2} E$ in Fig.~\ref{fig:1}.
However, the behaviour  of  the  curves  $\tilde{Q}_{\chi}(\tilde{\omega})$ and
$\tilde{E}(\tilde{\omega})$ is different in the neighbourhood of $\tilde{\omega
} = 1$.
Indeed, Figs.~\ref{fig:2} and  \ref{fig:3}  show  the  curves $\tilde{Q}_{\chi}
(\tilde{\omega})$   in   neighbourhoods  of  $\tilde{\omega} = 1$ for different
values of the gauge coupling constants.
The curves in  Fig.~\ref{fig:2}  correspond  to  the  same  values of the gauge
coupling constant as  in  Fig.~\ref{fig:1},  while  those  in  Fig.~\ref{fig:3}
correspond to larger values of the gauge coupling constants.
We see that unlike  the  curves $\tilde{E}(\tilde{\omega})$ in  the  subplot in
Fig.~\ref{fig:1}, the curves $\tilde{Q}_{\chi}(\tilde{\omega})$ end at the same
point.
It follows that the  limiting  thick-wall  value $\tilde{Q}_{\chi}(1)$ does not
depend on the  gauge coupling constants $\tilde{e}$ and $\tilde{q}$ separately.
Instead, it  depends  only  on  their  ratio  $\tau  =  \tilde{q}/\tilde{e}$ in
accordance with the conclusion of Sect.~\ref{sec:IV}.

\begin{figure}[tbp]
\includegraphics[width=0.5\textwidth]{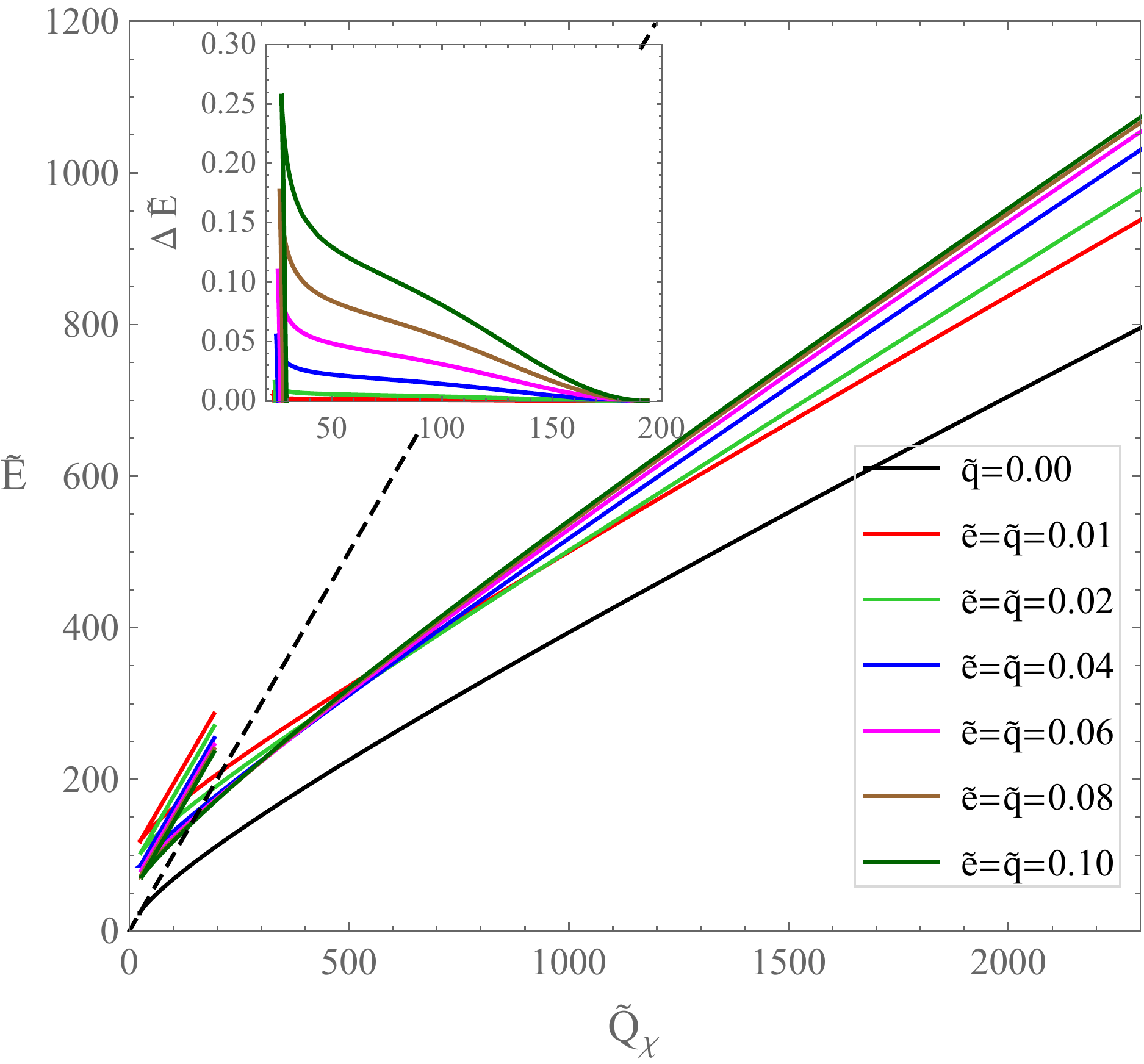}
\caption{Dependence of the energy $\tilde{E}$  of  the  vortex-Q-ball system on
the Noether  charge  $\tilde{Q}_{\chi}$  for  different  values  of  the  gauge
coupling constants.     The  straight  dashed  line  corresponds to  the linear
dependence $\tilde{E} = \tilde{Q}_{\chi}$. The subplot presents  the dependence
of the difference $\Delta\tilde{E}=\tilde{E}-\tilde{Q}_{\chi}-\tilde{E}_{\text{v
}}$ on $\tilde{Q}_{\chi}$ in the neighbourhood of cuspidal points
}
\label{fig:4}
\end{figure}

Figures~\ref{fig:2} and \ref{fig:3}  show  that the form of the curves $\tilde{
Q}_{\chi}(\tilde{\omega})$  changed  with  the increase in  the  gauge coupling
constants.
In   Fig.~\ref{fig:2},   the   curves  $\tilde{Q}_{\chi}( \tilde{\omega})$  are
$s$-shaped similar to the curves $\tilde{E}(\tilde{\omega})$  in the subplot in
Fig.~\ref{fig:1}.
At the same time, it  follows  from  Fig.~\ref{fig:3}  that  by  increasing the
gauge coupling constants,  the  curves $\tilde{Q}_{\chi}(\tilde{\omega})$ cease
to be $s$-shaped,  and   the  turning   points  of $s$-shaped $\tilde{Q}_{\chi}
(\tilde{\omega})$ curves  turn  into a single inflection point of monotonically
increasing $\tilde{Q}_{\chi}( \tilde{\omega})$ curves.

Figure~\ref{fig:4} presents the energy $\tilde{E}$  of the vortex-Q-ball system
as a function of its Noether charge $\tilde{Q}_{\chi}$ for  the  same values of
the gauge coupling constants as in Fig.~\ref{fig:1}.
We see that all the curves $\tilde{E}(\tilde{Q}_{\chi})$ that correspond to the
vortex-Q-ball system have one  cuspidal  point,  whereas  there is  no cuspidal
point  on  the  curve  $\tilde{E}(\tilde{Q}_{\chi})$  for  the  two-dimensional
nongauged Q-ball.
The next  characteristic  feature  is  that  the $\tilde{Q}_{\chi}$-coordinates
for the rightmost points of the upper branches of the curves $\tilde{E}(\tilde{
Q}_{\chi})$ coincide.
Of course,  this  is  a  consequence  of  the fact that  the value $\tilde{Q}_{
\chi}(1)$ depends on $\tilde{e}$ and $\tilde{q}$ only through the ratio $\tau =
\tilde{q}/\tilde{e}$,  which  is  the  same  for all  vortex-Q-ball  curves  in
Fig.~\ref{fig:4}.   

\begin{figure}[tbp]
\includegraphics[width=0.5\textwidth]{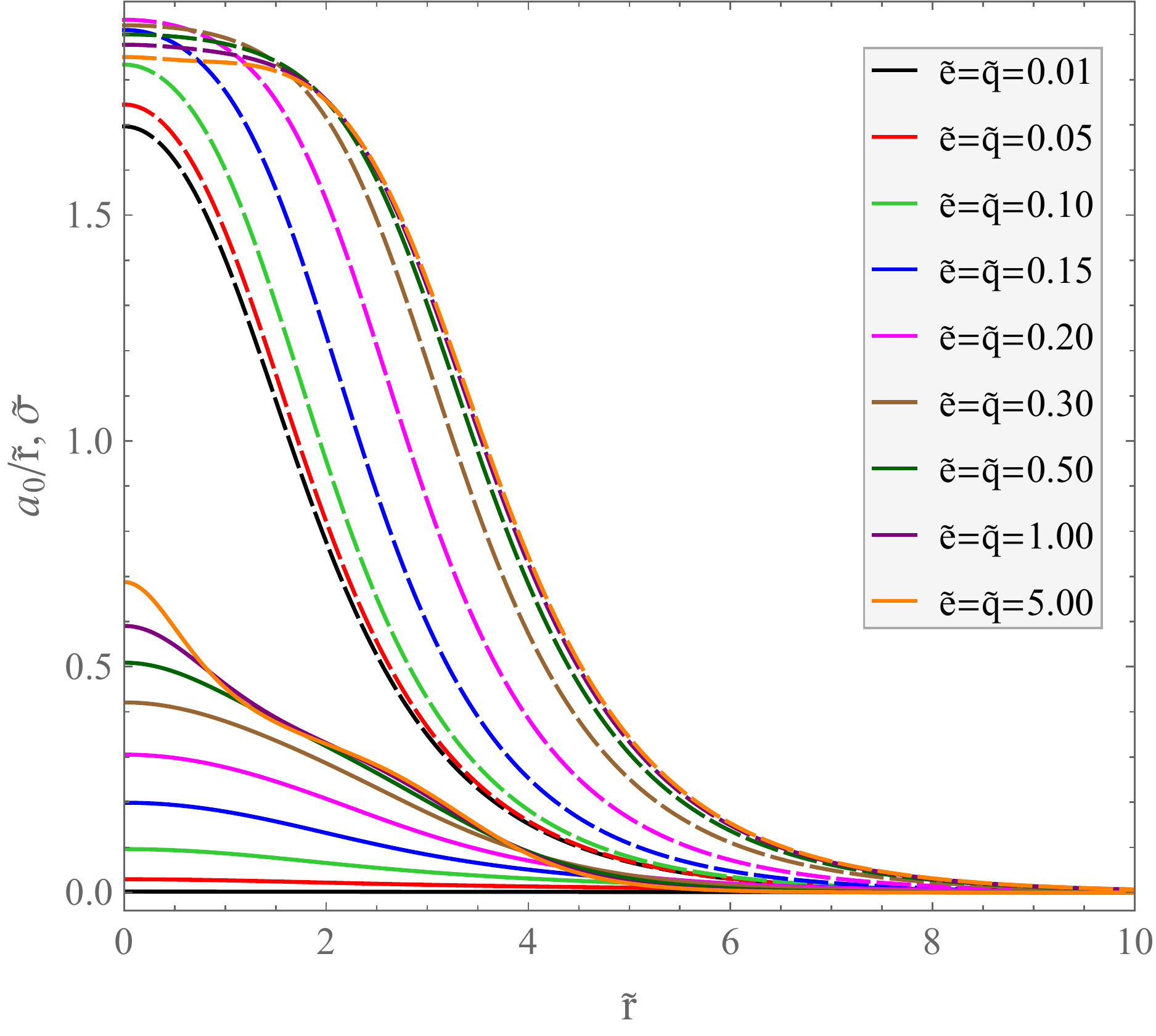}
\caption{Profile functions $a_{0}/\tilde{r}$ (solid curves) and $\tilde{\sigma}$
(dashed curves) of the vortex-Q-ball  system  for different values of the gauge
coupling constants. The curves correspond to parameters $\tilde{\omega} = 0.7$,
$\tilde{m} = 1$, $\tilde{g} = 1$, $\tilde{h} = 0.2$, $\tilde{\lambda} = 0.125$,
$\tilde{v} = 2$, $N = 1$, and $K = 0$}
\label{fig:5}
\end{figure}

In Fig.~\ref{fig:4}, the  dashed  straight  line  corresponds to the plane-wave
field configuration of the complex scalar field $\chi$.
We see that the energy  of  the two-dimensional nongauged $Q$-ball is less than
that of the  plane-wave  field  configuration, except for the point of contact,
at which they are equal.
Hence, the  two-dimensional  nongauged  $Q$-ball  is  stable against decay into
massive scalar $\chi$-bosons.
It follows from Fig.~\ref{fig:4} that  except for the neighbourhood of cuspidal
points, the energy of  the  vortex-Q-ball  system  is  less  than  that  of the
plane-wave  field  configuration.
Hence, the Q-ball component of the vortex-Q-ball system is stable against decay
into massive scalar $\chi$-bosons provided that $\tilde{E} < \tilde{Q}_{\chi}$.

\begin{figure}[tbp]
\includegraphics[width=0.5\textwidth]{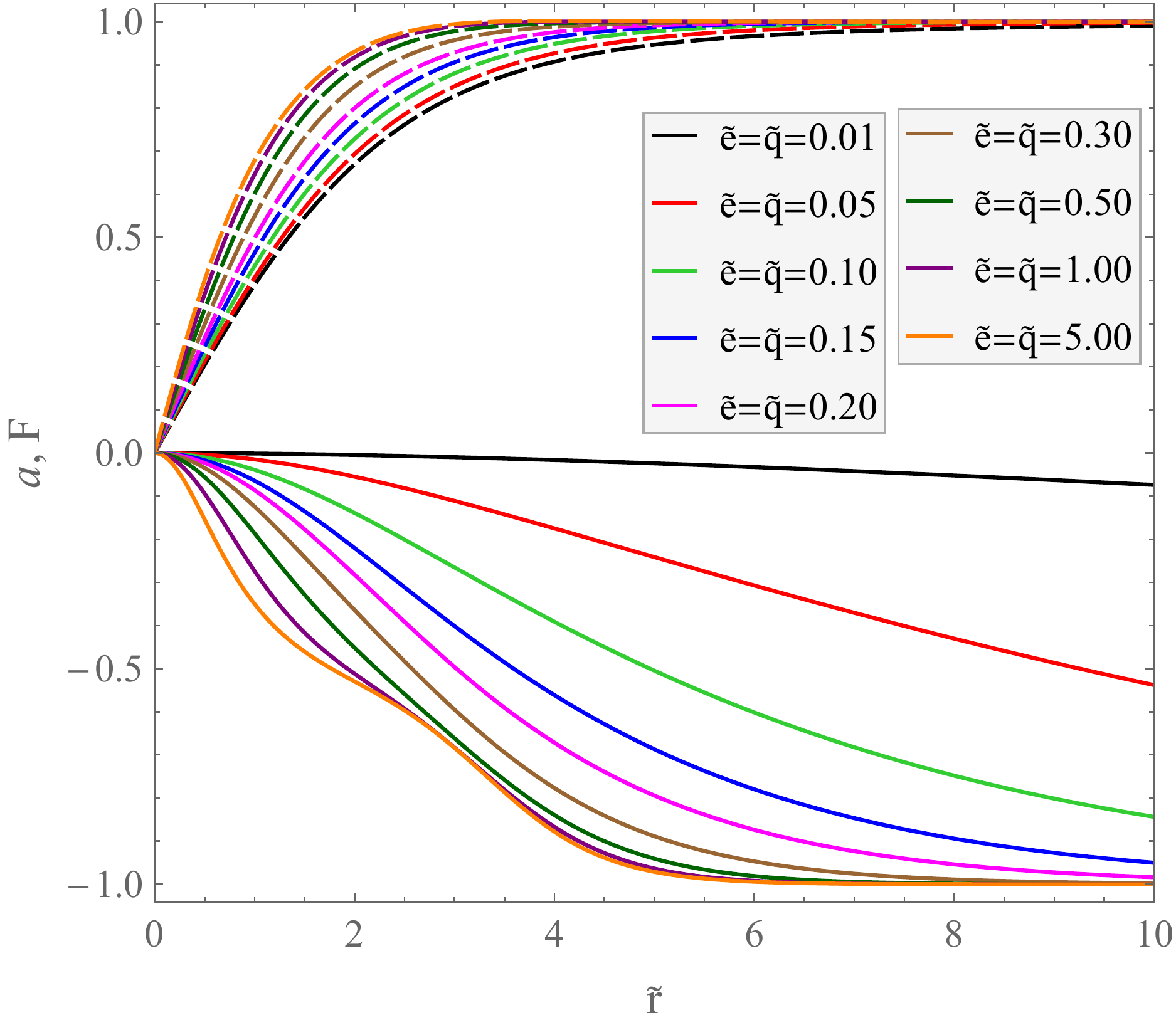}
\caption{Profile functions  $a$  (solid curves)  and $F$ (dashed curves) of the
vortex-Q-ball system  for different values of the gauge coupling constants. The
parameters of the vortex-Q-ball  system  are  the  same as in Fig.~\ref{fig:5}}
\label{fig:6}
\end{figure}

To  ascertain  the  possibility  of  decay  of  the  Q-ball  component  of  the
vortex-Q-ball system,  we  introduce  the  value $\Delta\tilde{E} = \tilde{E} -
\tilde{Q}_{\chi} - \tilde{E}_{\text{v}}$,  where  $\tilde{E}_{\text{v}}$ is the
energy of the vortex component of the system.
We define $\tilde{E}_{\text{v}}$ as  the  energy  of  the ANO vortex at a given
value of the gauge coupling constant $\tilde{e}$.
The curves $\tilde{E}_{\text{v}}(\tilde{Q}_{\chi})$ are presented in the subplot
in Fig.~\ref{fig:4}.
It is obvious that decay of  the  $Q$-ball component into scalar  $\chi$-bosons
is possible only if $\Delta\tilde{E}$ is positive.
It follows from the  subplot  in  Fig.~\ref{fig:4}  that for all gauge coupling
constants, $\Delta\tilde{E}$  is  positive  on  the  upper  branches  of curves
$\tilde{E}(\tilde{Q}_{\chi})$.
Furthermore, $\Delta\tilde{E}$  is  also  positive  on  the  lower  branches of
$\tilde{E}(\tilde{Q}_{\chi})$  curves   in  neighbourhoods  of  their  cuspidal
points.
Hence, the vortex-Q-ball  system  is  unstable  in  the  area  presented in the
subplot in Fig.~\ref{fig:4}.
The instability, however, can be either classical  (the presence of one or more
unstable modes  in   the   functional  neighbourhood  of the soliton system) or
quantum-mechanical (the possibility of quantum  tunneling of the soliton system
to another state).
It was shown in Refs.~\cite{lee_pang,fried_lee} that  the  appearance of a cusp
on the energy-Noether charge curve indicates the onset of a mode of instability.
Hence, the Q-ball components of the soliton systems lying on the upper branches
of the $\tilde{E}(\tilde{Q}_{\chi})$ curves are classically unstable.

Next we present the ansatz  functions of the vortex-Q-ball system for different
values of gauge coupling constants.
Figure~\ref{fig:5} presents the   ansatz functions $a_{0}(\tilde{r})/\tilde{r}$
and  $\tilde{\sigma}(\tilde{r})$, and  Figure~\ref{fig:6}  presents  the ansatz
functions $a(\tilde{r})$ and $F(\tilde{r})$.
It follows from Fig.~\ref{fig:5} that $\left. a_{0}\left(\tilde{r}\right)/\tilde{
r}\right\vert_{\tilde{r}= 0}$ increases monotonically and $a_{0}\left(\tilde{r}
\right)/\tilde{r}$  tends  to  limiting   form (\ref{IV:5a}) as $\tilde{e}$ and
$\tilde{q}$ increase.  
In particular, $\left. a_{0}\left(\tilde{r}\right)/\tilde{r}\right\vert_{\tilde{
r}= 0}$ reaches the maximum limit value $\tilde{\omega}/\tau=0.7$ as $\tilde{e}
= \tilde{q}  \rightarrow  \infty$, which is consistent with Eqs.~(\ref{III:24})
and (\ref{IV:5a}).
The ansatz  function $\tilde{\sigma}(\tilde{r})$  also tends to a limiting form
as the gauge coupling constants increase indefinitely.
Figure~\ref{fig:6}  shows  that  similar  to  $a_{0}(\tilde{r})/\tilde{r}$, the
ansatz function $a(\tilde{r})$  is  sensitive  to  the magnitude of $\tilde{e}$
and $\tilde{q}$.
In particular,   it  tends  to  zero  at any finite $\tilde{r}$ as $\tilde{e} =
\tilde{q}  \rightarrow  0$   and   tends  to  limiting  form  (\ref{IV:5b})  as
$\tilde{e} = \tilde{q}  \rightarrow  \infty$.
The dependence of $F(\tilde{r})$ on the the gauge coupling constants  is not as
strong as that of $a(r)$.
It follows from Fig.~\ref{fig:6} that $F(\tilde{r})$ tends to a  limiting  form
as  $\tilde{e}=\tilde{q} \rightarrow  \infty$ and to the global vortex solution
as  $\tilde{e} = \tilde{q} \rightarrow 0$.

\begin{figure}[tbp]
\includegraphics[width=0.5\textwidth]{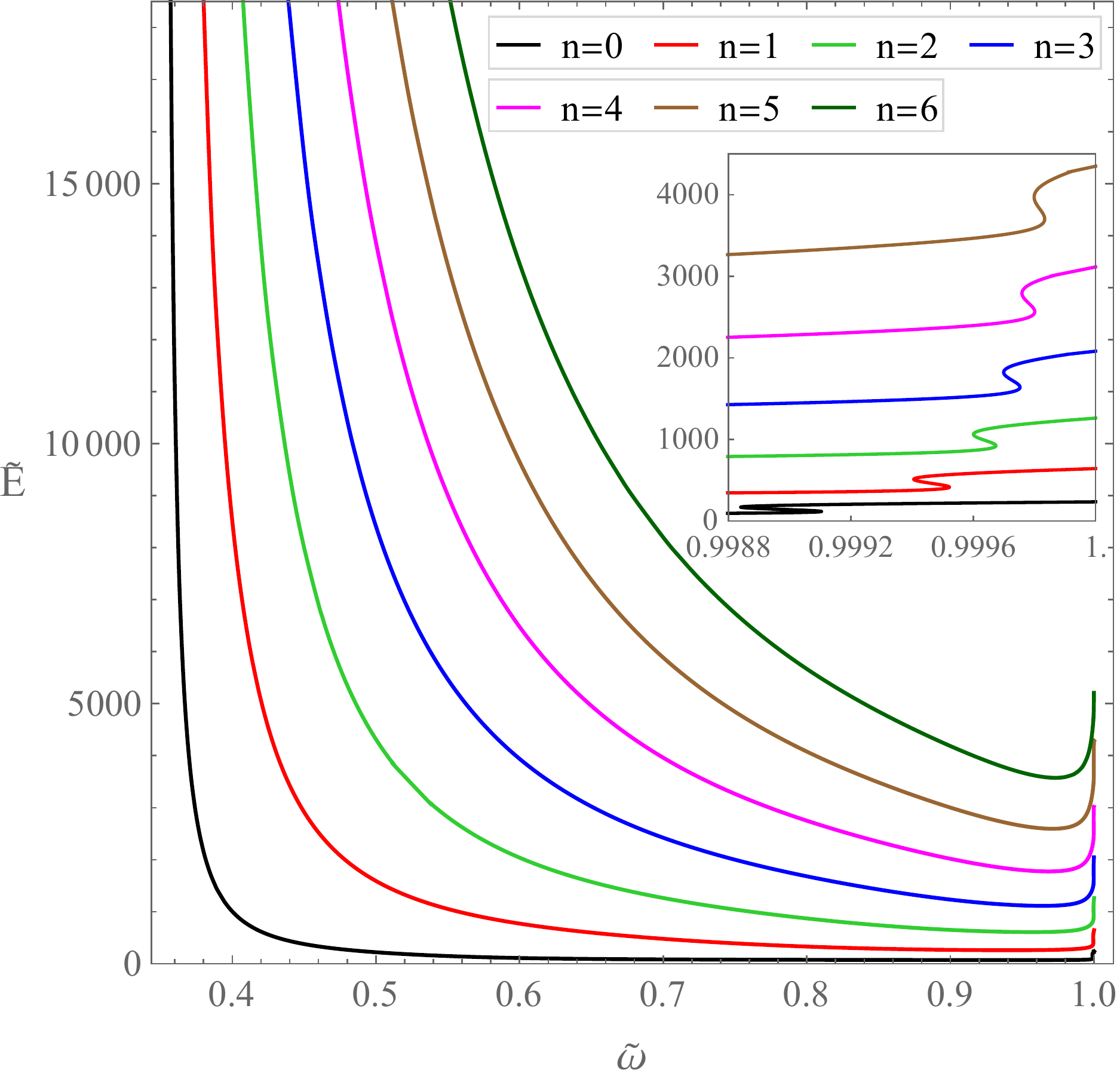}
\caption{Dependence of the energy $\tilde{E}$  of  the  vortex-Q-ball system on
the phase frequency $\tilde{\omega}$ for the first few radially excited states.
The curves correspond to parameters $\tilde{m} = 1$, $\tilde{g}= 1$, $\tilde{h}
= 0.2$,  $\tilde{\lambda} = 0.125$,  $\tilde{v} = 2$,  $\tilde{e} = \tilde{q} =
0.1$, $N = 1$, and $K = 0$}
\label{fig:7}
\end{figure}

Like  two  and  three-dimensional nongauged Q-balls, radially excited states in
the vortex-Q-ball system exist.
In Fig.~\ref{fig:7}, we can see the  curves $\tilde{E}(\tilde{\omega})$ for the
unexcited ($n = 0$)  and  the  first  six ($n = 1, \ldots, 6$) radially excited
states of the vortex-Q-ball system.
As in the previous (unexcited) case, the radially excited system passes into the
thin-wall regime as $\tilde{\omega} \rightarrow \tilde{\omega}_{\text{tn}}$ and
into the thick-wall regime as $\tilde{\omega} \rightarrow 1$.
Like curves  in  Fig.~\ref{fig:1},  the  curves  $\tilde{E}(\tilde{\omega})$ of
radially excited  vortex-Q-ball  systems  are  $s$-shaped  in  the  vicinity of
$\tilde{\omega} = 1$.
Next, we see that at fixed $\tilde{\omega}$, the energy of the the vortex-Q-ball
system increases with an increase in $n$.
It was found numerically that at fixed $\tilde{\omega}$, the energy and Noether
charge increase quadratically in $n$ starting with $n = 2$:
\begin{equation}
\tilde{E}\approx a  + b n^{2},\; \tilde{Q}_{\chi } \approx c + d n^{2},
                                                                    \label{V:2}
\end{equation}
where $a$ and $b$ are positive constants,  whereas  the  signs of constants $c$
and $d$ coincide with that of $\tilde{\omega}$.
The behaviour of $\tilde{Q}_{\chi}(\tilde{\omega})$  curves  is similar to that
of $\tilde{E}(\tilde{\omega})$ curves in Fig.~\ref{fig:7}.
In particular, the values  $\tilde{Q}_{\chi}(1)$  are  different  for different
$n$.

\begin{figure}[tbp]
\includegraphics[width=0.5\textwidth]{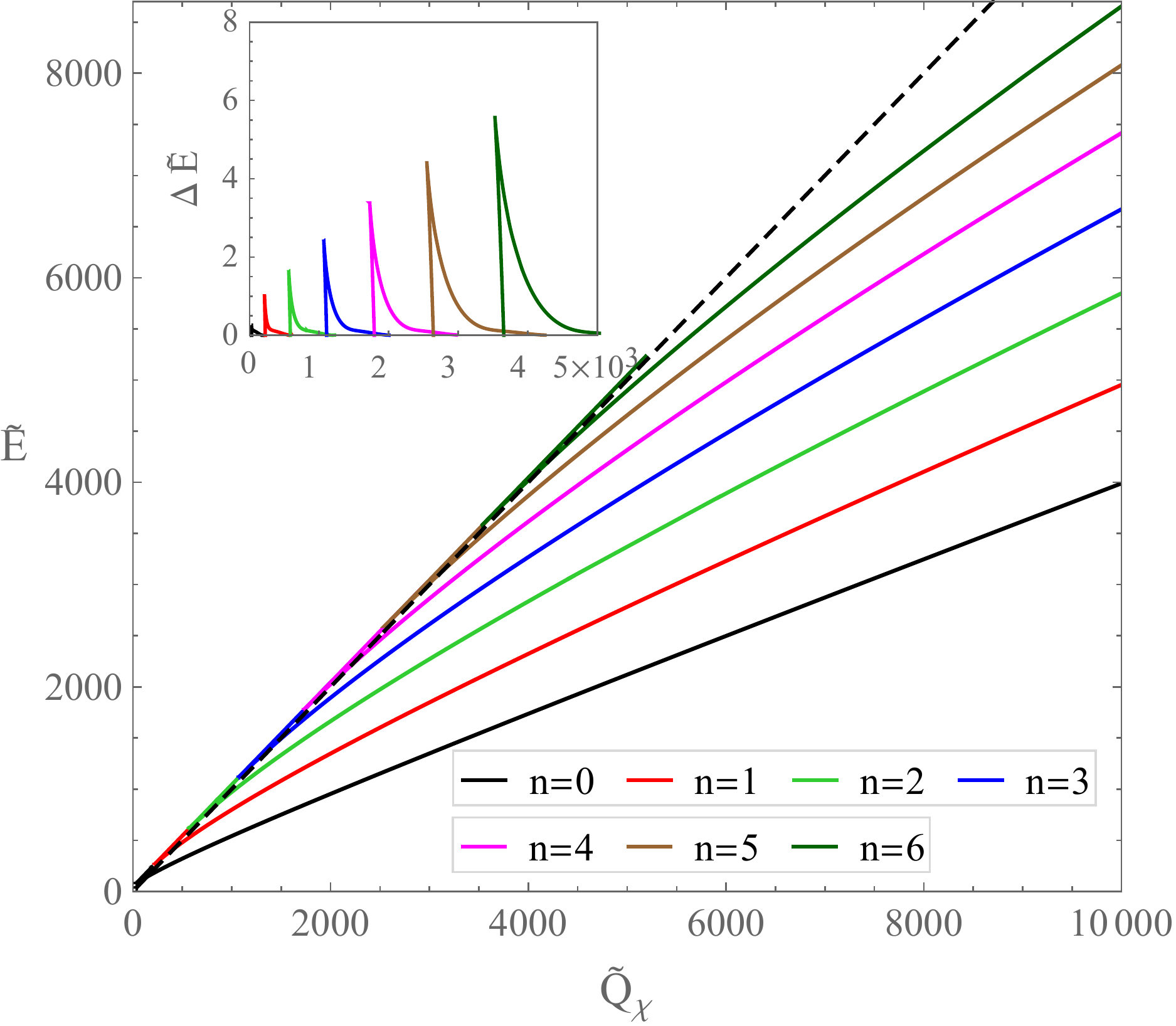}
\caption{Dependence of the energy $\tilde{E}$  of  the  vortex-Q-ball system on
the Noether charge $\tilde{Q}_{\chi}$ for the first few radially excited states.
The straight dashed line corresponds to the linear dependence $\tilde{E}=\tilde{
Q}_{\chi}$. The subplot presents the dependence of the difference $\Delta\tilde{
E}=\tilde{E}-\tilde{Q}_{\chi}  -\tilde{E}_{\text{v}}$  on $\tilde{Q}_{\chi}$ in
neighbourhoods of cuspidal points}
\label{fig:8}
\end{figure}

Figure~\ref{fig:8} presents  the  curves $\tilde{E}(\tilde{Q}_{\chi})$  for the
unexcited and the first six radially excited states of the vortex-Q-ball system.
All curves $\tilde{E}(\tilde{Q}_{\chi})$ have cuspidal points in which $\tilde{
E}$ and $\tilde{Q}_{\chi}$ reach minimum values.
Numerically, we found that starting with $n =2$, the minimum values of $\tilde{
Q}_{\chi}$ in the cusp points are well described  by  a  quadratic  dependence:
$Q_{\chi \text{c}}=a_{\text{c}}+b_{\text{c}}n^{2}$.
It follows that at  a  given $\tilde{Q}_{\chi}$, the number of radially excited
states of the vortex-Q-ball system does not exceed
\begin{equation}
n_{\max}=\left\lfloor b_{\text{c}}^{-1/2}\left(\tilde{Q}_{\chi } - a_{\text{c}}
\right)^{1/2}\right\rfloor,                                         \label{V:3}
\end{equation}
where $\lfloor x \rfloor$ denotes the floor  of  $x$ (the greatest integer less
than or equal to $x$).
We see that at large $\tilde{Q}_{\chi}$,  the number of radially excited states
rises $\propto \tilde{Q}_{\chi}^{1/2}$.

It follows  from  Fig.~\ref{fig:8}  that  at  a  given  $\tilde{Q}_{\chi}$, the
lower-branch energy $\tilde{E}$ increases with $n$.
Hence, radially excited vortex-Q-ball states lying on lower branches of $\tilde{
E}(\tilde{Q}_{\chi})$ curves are  unstable  with respect to the transition into
less excited states.
Furthermore, we can see from the  subplot  in  Fig.~\ref{fig:8} that the Q-ball
components of radially excited vortex-Q-ball  states  are unstable with respect
to decay into scalar $\chi$-bosons  in  neighbourhoods  of cuspidal points (the
upper branches and small parts of the lower branches  of $\tilde{E}(\tilde{Q}_{
\chi})$ curves).

\begin{figure}[tbp]
\includegraphics[width=0.5\textwidth]{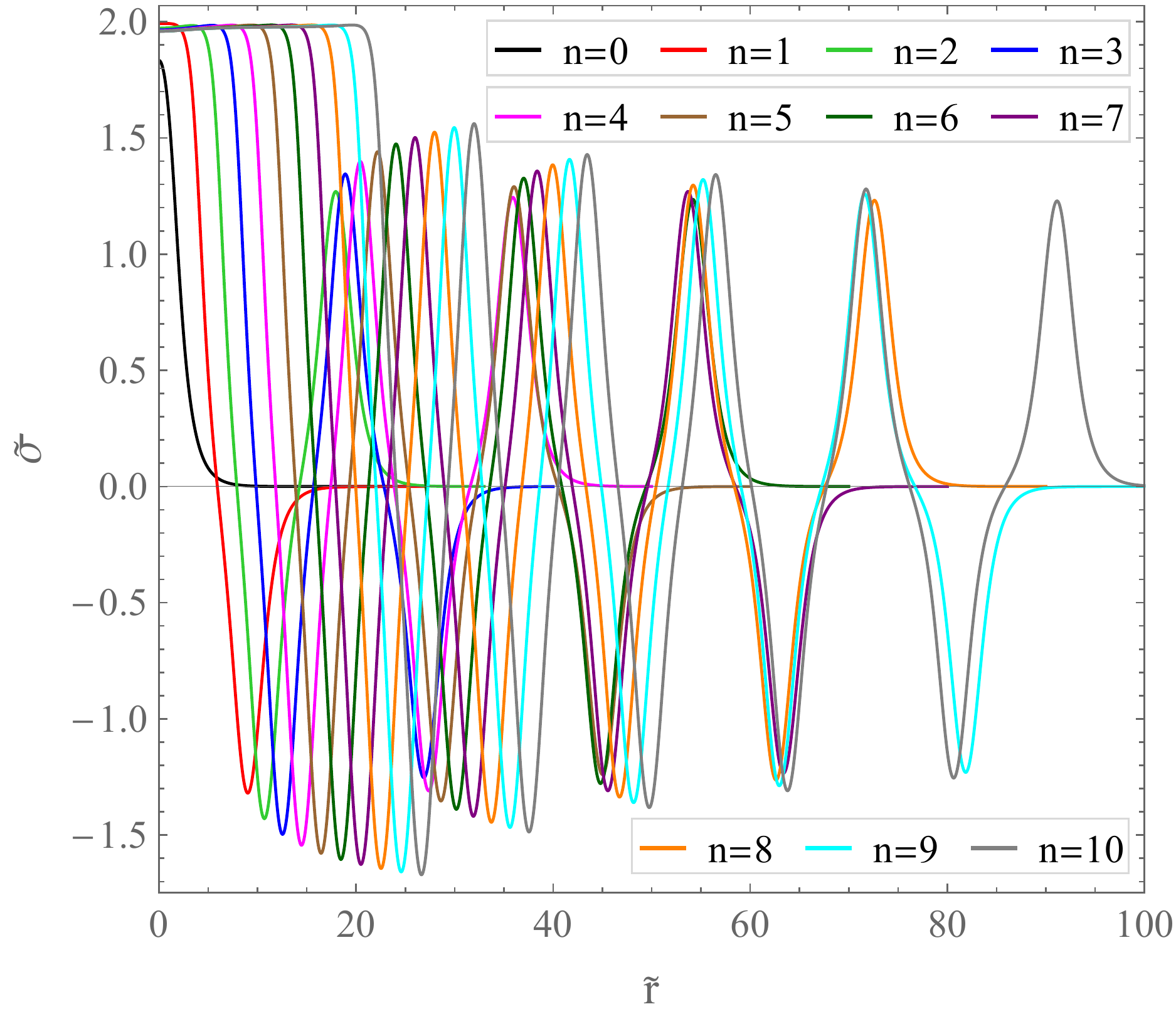}
\caption{Profile functions $\tilde{\sigma}$  for the first 10  radially excited
states of  the  vortex-Q-ball  system.  The  curves  correspond  to  parameters
$\tilde{\omega} = 0.7$,  $\tilde{m} = 1$,  $\tilde{g} = 1$,  $\tilde{h} = 0.2$,
$\tilde{\lambda} = 0.125$, $\tilde{v} = 2$, $\tilde{e} = \tilde{q} = 0.1$, $N =
1$, and $K = 0$}
\label{fig:9}
\end{figure}

\begin{figure}[tbp]
\includegraphics[width=0.5\textwidth]{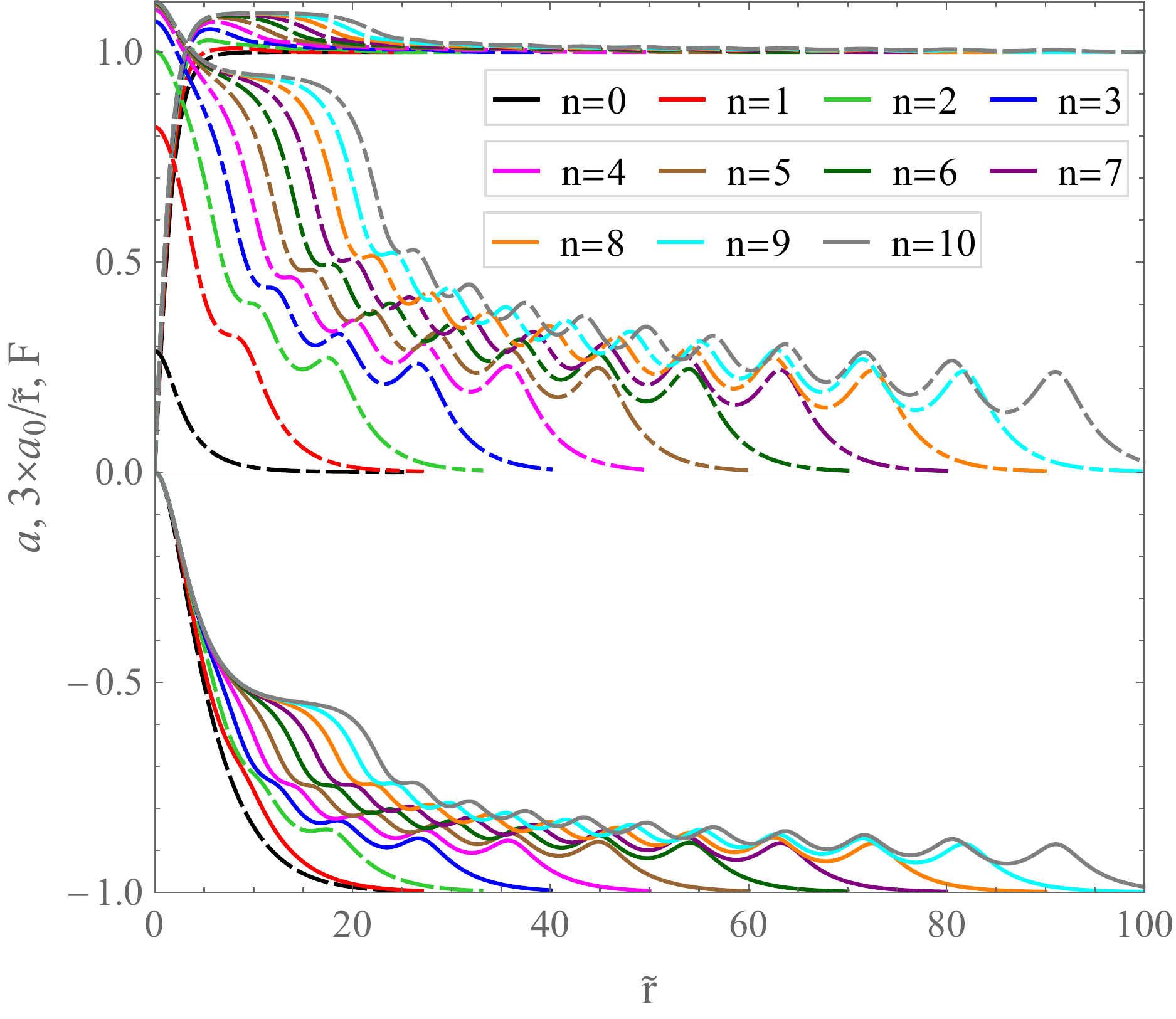}
\caption{Ansatz functions  $a$,  $3a_{0}/\tilde{r}$, and $F$  for the first  10
radially excited states of  the  vortex-Q-ball  system.   The parameters of the
vortex-Q-ball system are the same as in Fig.~\ref{fig:9}}
\label{fig:10}
\end{figure}

Figures~\ref{fig:9} and \ref{fig:10} present ansatz  functions for the first 10
radially excited states of the vortex-Q-ball system.
We see from Fig.~\ref{fig:9} that oscillations of the $\tilde{\sigma}(\tilde{r})$
curves are inharmonious.
In particular, amplitudes of peaks  and  valleys of $\tilde{\sigma}(\tilde{r})$
curves  are  decreased  with  the  increase  in  the  number  of  oscillations,
whereas the distances  between  adjacent  peaks  and  valleys of $\tilde{\sigma
}(\tilde{r})$ curves are increased.
As  seen  in  Fig.~\ref{fig:10}, the  ansatz  functions  $\tilde{a}_{0}(\tilde{
r})/\tilde{r}$ and $a(\tilde{r})$ also oscillate.
In  particular,  the  positions  of  local  maxima  of  $\tilde{a}_{0}(\tilde{r
})/\tilde{r}$ and $a(\tilde{r})$ approximately coincide with those of peaks and
valleys of $\tilde{\sigma}(\tilde{r})$, whereas the positions  of  local minima
of $\tilde{a}_{0}(\tilde{r})/\tilde{r}$  and  $a(\tilde{r})$  are approximately
the same as those of nodes (zeros) of $\tilde{\sigma}(\tilde{r})$.

It was shown in Ref.~\cite{loginov_prd_102} that the number of radially excited
states of the three-dimensional gauged Q-ball is finite.
At the same time, we found  no  indication  of  the finiteness of the number of
radially excited states for the vortex-Q-ball system.
Of course, the reason for this difference is that the  three-dimensional gauged
Q-ball possesses an electrical charge, whereas the two-dimensional vortex-Q-ball
system is electrically neutral.
As a result, the electric charge density of the three-dimensional gauged Q-ball
is always either  positive  or  negative,  whereas  that  for the vortex-Q-ball
system is alternating.
This difference is apparent in the behaviour of the ansatz function $\Omega(r)$
defined in  Eq.~(\ref{III:22}).
Indeed,  it  is   shown   in   Ref.~\cite{klee}  that  in   the   case  of  the
three-dimensional gauged Q-ball, $\Omega(r)$ is a bounded ($0<\Omega(r)<\omega$)
and monotonically increasing function of $r$.
For the vortex-Q-ball system, the  ansatz function $\Omega(r)$ is also  bounded
in the interval $(0, \omega)$ as it is seen from Eq.~(\ref{III:23}).
However, $\Omega(r)$ need not be monotonic in this case.
Indeed, the oscillating behaviour of $\tilde{a}_{0}(\tilde{r})/\tilde{r}$ shown
in Fig.~\ref{fig:10} results in the oscillating behaviour of $\Omega(\tilde{r})
= \omega - \tau \tilde{a}_{0}(\tilde{r})/\tilde{r}$.
The monotonic increase of $\Omega(r)$ leads to the restriction of the number of
radially  excited  states  for   the   three-dimensional gauged Q-ball, because
in this case, $\Omega(r)$ reaches its  maximum  value  $\omega$  for  a  finite
number of oscillations of $\sigma(r)$.
In contrast, the  nonmonotonic  oscillating  behaviour  of $\Omega(r)$ makes it
possible  for  vortex-Q-ball  systems   with  an arbitrarily  large  number  of
oscillations (and hence nodes)  of  $\sigma(r)$  to  exist;  thus, there are no
restrictions on the  number  of  radially   excited states of the vortex-Q-ball
system.

\begin{figure}[tbp]
\includegraphics[width=0.5\textwidth]{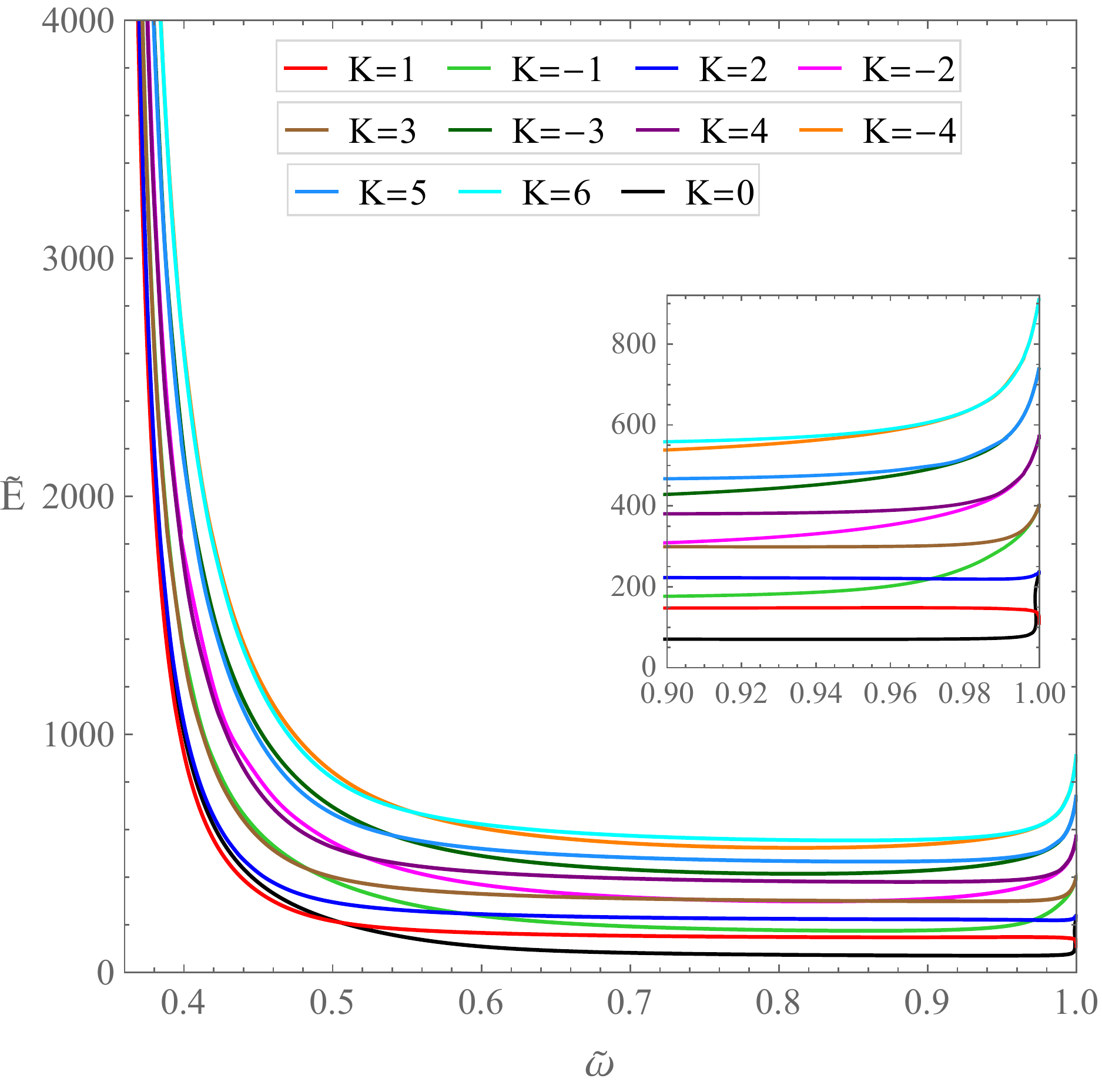}
\caption{Dependence of the energy $\tilde{E}$  of  the  vortex-Q-ball system on
the phase frequency $\tilde{\omega}$  for  the  first  few  azimuthally excited
states.    The curves correspond to parameters  $\tilde{m} = 1$, $\tilde{g}=1$,
$\tilde{h} = 0.2$,  $\tilde{\lambda} = 0.125$,   $\tilde{v} = 2$,  $\tilde{e} =
\tilde{q} = 0.1$, and $N = 1$}
\label{fig:11}
\end{figure}

Now we turn  to  a  description  of the  vortex-Q-ball's  states  with  nonzero
integer-valued parameter $K$.
Figure~\ref{fig:11} shows the curves  $\tilde{E}(\tilde{\omega})$ for the first
few states with nonzero $K$.
It follows from  Fig.~\ref{fig:11}  that  any  two  $\tilde{E}(\tilde{\omega})$
curves whose parameters $K'$ and $K''$ satisfy the condition $K' +K'' = 2$ tend
to the same limit in the thick-wall regime when $\tilde{\omega} \rightarrow 1$.
A similar statement is valid for the $\tilde{Q}_{\chi}(\tilde{\omega})$ curves.
This facts can be explained as follows.
In the thick-wall regime,  the  limiting energy  of the vortex-Q-ball system is
written  as  $E_{\text{tk}} = F_{0} + 2 m^{2}\bar{F}_{2}$, where $F_{0}$ is the
energy of the ANO vortex  with  a  given  $e$,  and  $2m^{2}\bar{F}_{2}$ is the
energy of the Q-ball component.
It follows from  Eq.~(\ref{IV:11e}) that  $\bar{F}_{2}$  is expressed solely in
terms of the rescaled ansatz function $\bar{\sigma}$.
Next, the  rescaled   ansatz  function  $\bar{\sigma}$  satisfies  differential
equation (\ref{IV:15}).
We see that Eq.~(\ref{IV:15}) depends on the  integer-valued parameters $K$ and
$N$ only through the combination $\varkappa = (K - \tau N)^{2}$.
It follows that the parameters $K'$ and $K''$, which satisfy the relation $K' +
K'' = 2 \tau N$,  lead to the same value of $\varkappa$.
Because both $K'$  and  $K''$  are integers, the value of $2 \tau N$ is also an
integer, thus the parameter  $\tau  =  e/q$  is  an  integer  or  half-integer.
If  $2 \tau  N$  is  a   noninteger,  then  $\tilde{E}(\tilde{\omega})$  curves
corresponding  to  different  $K$  will  not tend to the same limit as $\tilde{
\omega} \rightarrow 1$.
In our case, the parameters $N$ and $\tau$ are equal to one, so $2 \tau N = 2$.
Thus, differential equation (\ref{IV:15})  will  have the same form for any two
vortex-Q-ball systems whose parameters satisfy the condition $K' + K'' = 2$.

\begin{figure}[tbp]
\includegraphics[width=0.5\textwidth]{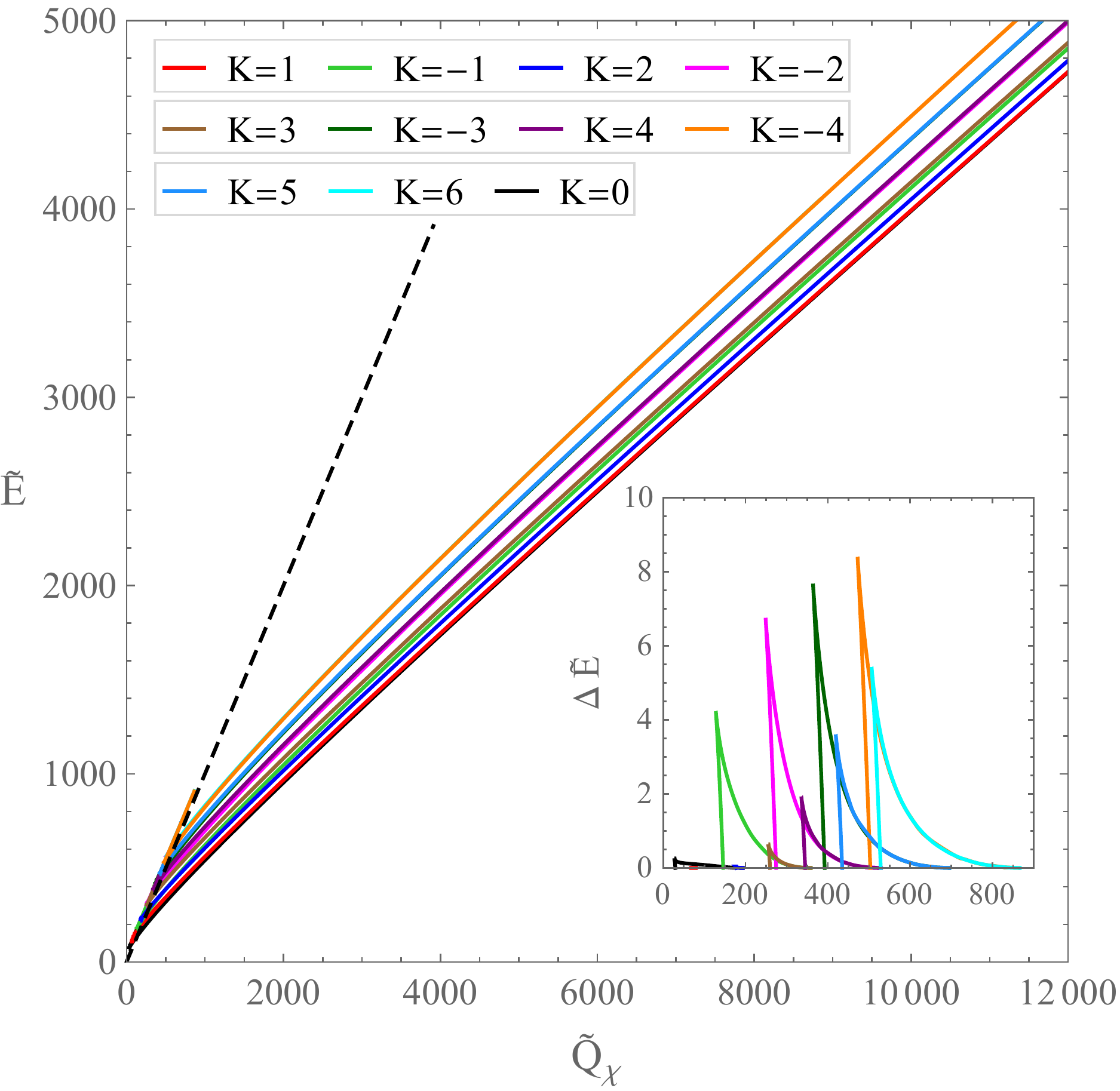}
\caption{Dependence of the energy $\tilde{E}$  of  the  vortex-Q-ball system on
the Noether charge $\tilde{Q}_{\chi}$  for  the  first  few azimuthally excited
states. The straight dashed line corresponds to the linear dependence $\tilde{E}
= \tilde{Q}_{\chi}$.  The subplot presents  the  dependence  of  the difference
$\Delta\tilde{E} = \tilde{E}-\tilde{Q}_{\chi}-\tilde{E}_{\text{v}}$ on $\tilde{
Q}_{\chi}$ in neighbourhoods of cuspidal points}
\label{fig:12}
\end{figure}

Eq.~(\ref{IV:15}) is valid for all  $\bar{r}$   except those in the interval $0
\leq \bar{r}\lesssim \Delta_{\omega}m_{A, \phi}^{-1}$ whose width tends to zero
in the thick-wall regime.
This is because the rescaled ansatz functions $\bar{a}(\bar{r})\equiv a(\Delta_{
\omega}^{-1} \bar{r})$   and   $\bar{F}(\bar{r}) \equiv  F(\Delta_{\omega}^{-1}
\bar{r})$ are different  from  their  limiting  values  in  this  infinitesimal
interval of $\bar{r}$.
Next,  the  ansatz   functions  $\bar{\sigma}_{K'}(\bar{r})$  and  $\bar{\sigma
}_{K''}(\bar{r})$ (the dependence of $\bar{\sigma}$ on $K$ is shown explicitly)
satisfy  the  same  boundary  conditions:  $\bar{\sigma}(0)  =  0$  and  $\bar{
\sigma}(\infty) = 0$.
The only exception is in the case $K = 0$ for which the left boundary condition
is  $\bar{\sigma}'(0) = 0$.
However, it was found numerically that in this case, $\bar{\sigma}(0)\rightarrow
0$ as $\Delta_{\omega} \rightarrow 0$, and  the  left  boundary  condition  for
$K'=0$  becomes essentially the same as that for the complementary case $K''=2$.
It follows that  the  ansatz  functions $\bar{\sigma}_{K'}(\bar{r})$ and $\bar{
\sigma}_{K''}(\bar{r})$ that  satisfy  the  condition $K'+ K'' = 2$ tend to the
same limit in the thick-wall regime  and  so  do  the corresponding functionals
$\bar{F}_{2K'}$ and $\bar{F}_{2K''}$.
Hence,  the  energies (Noether charges)  of  the two vortex-Q-ball systems with
$K'+ K'' = 2$  tend  to the same value in the thick-wall regime.
Note that there is no complementary vortex-Q-ball system for the system with $K
= 1$.
Indeed, in this case the parameters $K'$ and $K''$ are the same: $K^{\prime}=1,
K^{\prime \prime } = 1 \Rightarrow K^{\prime } + K^{\prime \prime} = 2$.

It follows from Eq.~(\ref{III:29})  that  the  two  vortex-Q-ball  systems with
the same Noether charges and with integer  parameters $K'$ and $K''$  such that
$K' + K'' = 2\tau N$ possess the opposite angular momenta.
Hence, the angular momenta  of  two  vortex-Q-ball  systems with $K' + K'' = 2$
tend to the  opposite  values  in  the  thick-wall  regime, whereas the angular
momentum of the state with $K = 1$ vanishes.

Figure~\ref{fig:12}  presents  the  $\tilde{E}({\tilde{Q}_{\chi}})$  curves for
the same $K$ as in Fig.~\ref{fig:11}.
We see that just as in Figs.~\ref{fig:4} and \ref{fig:8}, the  curves $\tilde{E
}({\tilde{Q}_{\chi}})$ have cuspidal points.
The only exception is the $\tilde{E}({\tilde{Q}_{\chi}})$ curve for $K = 1$; in
this case, the absence of a cuspidal point follows from the monotonicity of the
corresponding $\tilde{E}({\tilde{\omega}})$ curve in Fig.~\ref{fig:11}.
The conservation of the Noether  charge  and  the angular momentum leads to the
conclusion that the parts of $\tilde{E}({\tilde{Q}_{\chi}})$ curves lying below
the line  $\tilde{E} = \tilde{E}_{\text{v}}  + \tilde{Q}_{\chi}$  correspond to
stable vortex-Q-ball systems.
In the subplot in Fig.~\ref{fig:12}, we can see the parts of $\tilde{E}({\tilde{
Q}_{\chi}})$ curves that lie above the line $\tilde{E} = \tilde{E}_{\text{v}} +
\tilde{Q}_{\chi}$  and  thus  correspond  to  unstable  vortex-Q-ball  systems.

It follows from Fig.~\ref{fig:12} that the curves $\tilde{E}_{K^{'}}(\tilde{Q_{
\chi}})$ and $\tilde{E}_{K^{''}}(\tilde{Q_{\chi}})$ with $K' + K'' = 2 \tau N =
2$ approximately coincide  starting with $\left(K^{\prime }, K^{\prime \prime }
\right) =(4,-2)$.
Hence, the  energies  of  the  two  vortex-Q-ball  systems  that correspond  to
coincident curves  $\tilde{E}_{K^{'}}(\tilde{Q_{\chi}})$ and $\tilde{E}_{K^{''}
}(\tilde{Q_{\chi}})$ are approximately the same at  given  $Q_{\chi}$.
Note that the parameter $K-\tau N$  has opposite values for the curves $\tilde{
E}_{K^{'}}(\tilde{Q_{\chi}})$  and $\tilde{E}_{K^{''}}(\tilde{Q_{\chi}})$  with
$K' + K'' = 2 \tau N$.
Hence, the  angular  momenta $\tilde{J}_{K^{'}}(\tilde{Q_{\chi}})$ and $\tilde{
J}_{K^{''}}( \tilde{Q_{\chi}} )$   are   opposite    as    it    follows   from
Eq.~(\ref{III:29}).
It follows from  Eqs.~(\ref{III:31a})--(\ref{III:31c}) that  under  the reverse
$(K, N) \rightarrow (-K, -N)$, the angular momentum of the vortex-Q-ball system
changes the sign, whereas the energy  and  the Noether charge remains the same.
The parameter $K-\tau N$  also  changes  the  sign  under  the  reverse $(K, N)
\rightarrow (-K, -N)$.
Hence, the functions  $E(N,K)$, $Q_{\chi}(N,K)$, and $J(N,K)$ in essence become
functions of one argument $K-\tau N$ when the  parameter $\left\vert K - \tau N
\right\vert \gtrsim 3$.

\begin{figure}[tbp]
\includegraphics[width=0.5\textwidth]{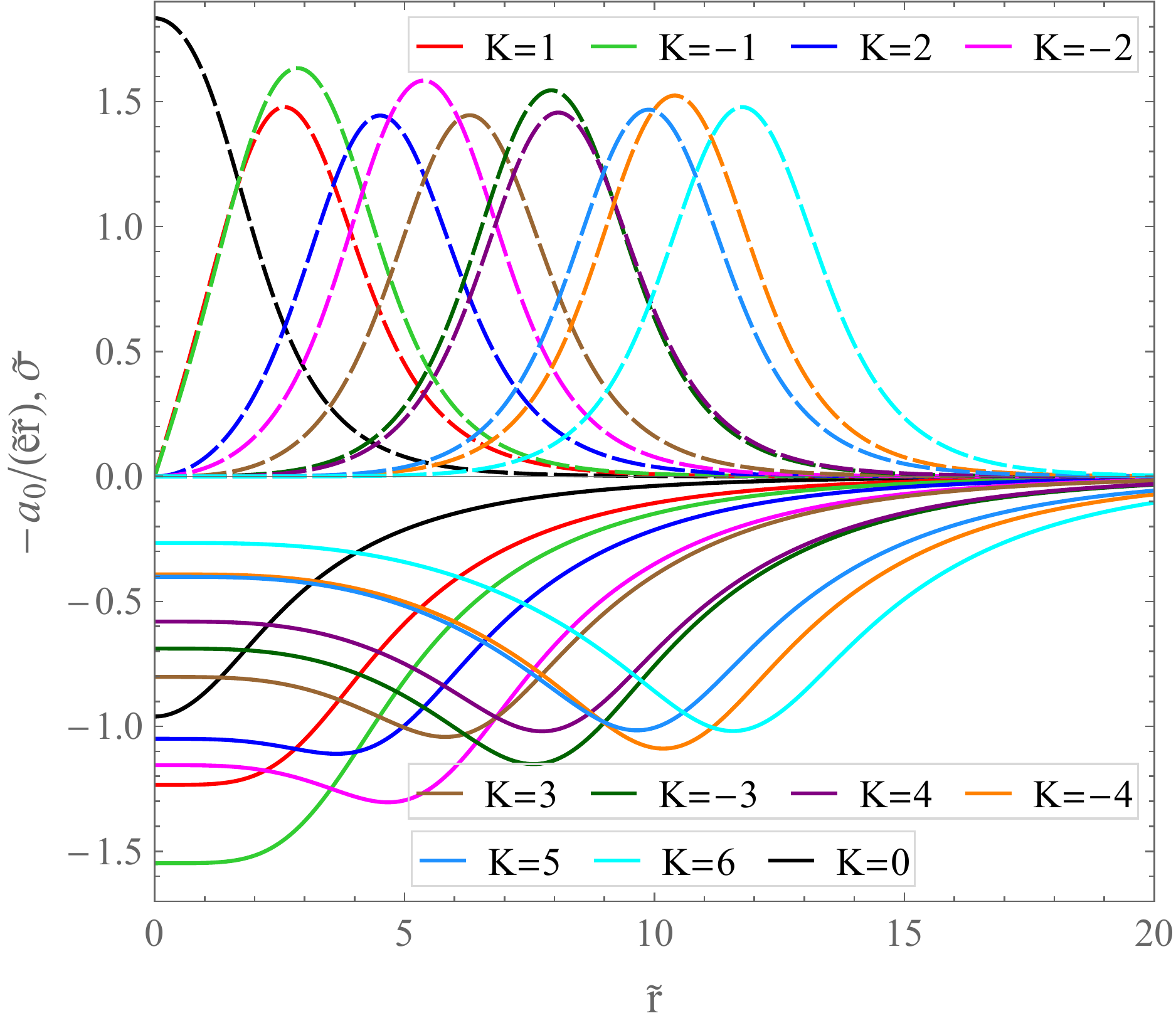}
\caption{Profile  functions  -$a_{0}/(\tilde{e}\tilde{r})$  (solid  curves) and
$\tilde{\sigma}$ (dashed curves) of the vortex-Q-ball  system for the first few
azimutally excited states. The curves correspond to parameters $\tilde{\omega}=
0.7$,  $\tilde{m} = 1$, $\tilde{g} = 1$, $\tilde{h} = 0.2$,  $\tilde{\lambda} =
0.125$,  $\tilde{v} = 2$,  $\tilde{e} = \tilde{q} = 0.1$, and $N = 1$}
\label{fig:13}
\end{figure}

\begin{figure}[tbp]
\includegraphics[width=0.5\textwidth]{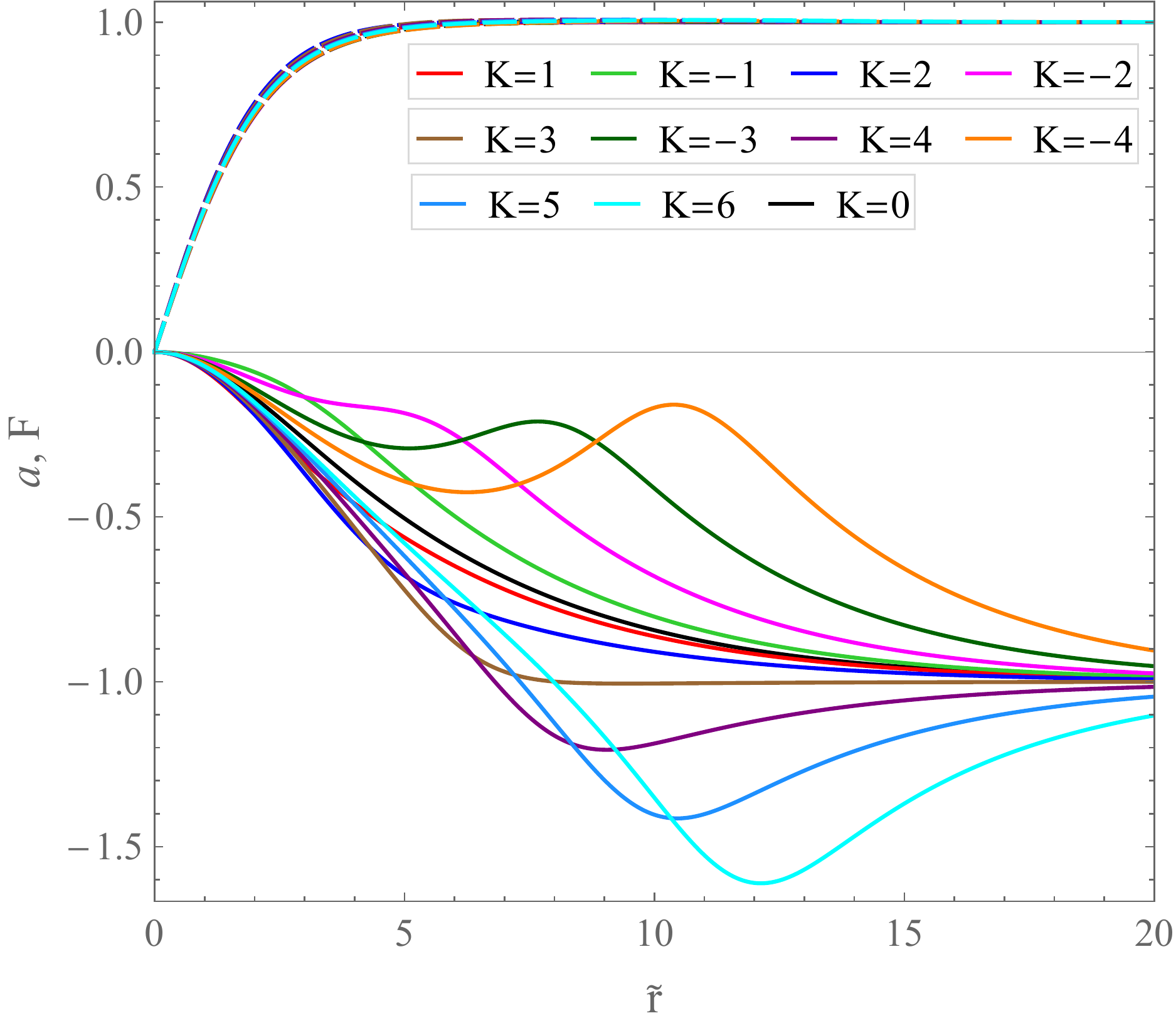}
\caption{Ansatz  functions  $a$  (solid curves)  and $F$ (dashed curves) of the
vortex-Q-ball  system  for  the  first  few  azimuthally  excited  states.  The
parameters of the vortex-Q-ball system are the same as in Fig.~\ref{fig:13}}
\label{fig:14}
\end{figure}

The reason for this is the presence of the  centrifugal  term $-r^{-2}\left(K +
\tau a\left( r\right) \right)^{2}\sigma \left( r \right)$ in Eq.~(\ref{III:6}).
The contribution  of  this  term  is  proportional  to  $r^{-2}\left(K - \tau N
\right)^{2}$ when $a(r)$ is in the vicinity of the limiting value of $-N$.
The growth of $\left\vert K - \tau N \right\vert$  leads  to  the growth of the
factor $\varkappa = \left(K - \tau N\right)^{2}$ in the  centrifugal term.
The growth of $\varkappa$ must be compensated, otherwise, the solution of mixed
boundary value problem (\ref{III:3})--(\ref{III:6})  and (\ref{III:8}) will not
exist.
Such  compensation  is  achieved  through  the  shift  of  the  ansatz function
$\sigma(r)$ towards higher values of $r$.
The shift results in reducing the factor $r^{-2}$ in the centrifugal term which
compensates for the growth of the factor $\varkappa$.
With increasing  $r$, it  is  convenient  to  pass  to  the new ansatz function
$\Delta a(r) = N + a(r)$.
Then the system (\ref{III:3})--(\ref{III:6}) will depend  on the parameters $K$
and $N$ only through the combination $K - \tau N$.

Figures~\ref{fig:13} and \ref{fig:14} show  the  ansatz  functions for the same
$K$ as in Fig.~\ref{fig:11}.
We see that  for  $K  \ne  0$, the ansatz functions $\tilde{\sigma}(\tilde{r})$
have a slightly asymmetric bell-shaped form.
Numerically,  we  found  that  the  radial  positions of maxima of the $\tilde{
\sigma}(\tilde{r})$ curves increase approximately  linearly  with $\left\vert K
- \tau N \right\vert$.
It follows from  Fig.~\ref{fig:13}  that  with  an  increase  in  $\left\vert K
- \tau N \right\vert$, the ansatz functions $a_{0}(\tilde{r})/(\tilde{e}\tilde{
r})$  also  form  maxima  whose  radial positions are approximately the same as
those of the corresponding $\tilde{\sigma}(\tilde{r})$ curves.
Next, we see from Fig.~\ref{fig:14} that  the  forms  of  $a(\tilde{r})$ curves
strongly depend on $K - \tau N$.
In particular, with an increase in $\left\vert K-\tau N\right\vert$, the minimum
of $a(\tilde{r})$ appears for positive  $K$  and the maximum (together with the
left adjacent  minimum) of $a(\tilde{r})$ appears for negative $K$.
The radial positions of these extremes of $a(\tilde{r})$ approximately coincide
with those of the maxima of the corresponding $\tilde{\sigma}(\tilde{r})$.
Such a difference in the behaviour of the $a(\tilde{r})$  curves  with positive
and negative $K$ is explained by the  fact  that  the  driving  term $-2 e q (K
- \tau N) \sigma(r)^{2}$  in  Eq.~(\ref{III:4}) has  the opposite sign in these
cases.
Finally, it follows from Fig.~\ref{fig:14} that the forms of the $F(\tilde{r})$
curves weakly depend on $K$.

\section{Conclusion}                                             \label{sec:VI}

In  the  present  paper,  we  continued  the study of the vortex-Q-ball systems
started in Ref.~\cite{loginov_plb_777}.
We investigated properties of the unexcited vortex-Q-ball systems  at different
values  of  gauge  coupling  constants  and  properties  of   radially  excited
vortex-Q-ball systems.
We also investigated properties of azimuthally  excited  vortex-Q-ball systems.
The vortex-Q-ball system is composed  of  a  vortex (topological soliton) and a
Q-ball   (nontopological   soliton),  thus  it  combines   properties  of  both
topological and nontopological solitons.
In  particular, the  vortex-Q-ball  possesses  a  quantized  magnetic  flux and
satisfies basic relation (\ref{III:1}) for nontopological solitons.

The  electromagnetic interaction  between  the  vortex  and  Q-ball  components
results in significant changes in their properties.
Firstly, the vortex and Q-ball components of  the  vortex-Q-ball system acquire
opposite electrical charges, so the  system  remains  electrically neutral as a
whole.
At  the  same  time,  neither  ANO  vortex  nor  two-dimensional Q-ball possess
electrical charge because in  gauge  models  without the Chern-Simons term, any
two-dimensional  electrically charged object will have infinite energy.
Secondly, the vortex-Q-ball system  possesses  nonzero angular momentum despite
the fact that both the  ANO  vortex  and  the two-dimensional axially symmetric
Q-ball have zero angular momenta.
Thirdly, the interaction  between  the  vortex and  Q-ball  components leads to
substantial  change  in  the  $E(\omega)$  and  $E(Q_{\chi})$  curves  for  the
vortex-Q-ball system in comparison with  that of  the two-dimensional nongauged
Q-ball.
In particular, in  the  case  of  the vortex-Q-ball system, some of $E(\omega)$
curves are $s$-shaped in the vicinity of  $\omega = m$  and  the  $E(Q_{\chi})$
curves have  cuspidal  points,   whereas   none   of   these  features hold for
the two-dimensional nongauged Q-ball.

There  exist   several   extreme   regimes   for   the   vortex-Q-ball  system.
We  investigated  four of them:  the  thick-wall  regime  in  which  the  phase
frequency $\omega$ tends to the  maximum  value,  the thin-wall regime in which
$\omega$ tends to the  minimum  value,  and  regimes  of  small and large gauge
coupling constants.
In particular, we  found  that  the  limiting  thick-wall  value of the Noether
charge depends on  the  gauge  coupling  constants $e$ and $q$ only through the
ratio $\tau = q/e$.
We also found that the limiting thick-wall energies of two vortex-Q-ball systems
whose azimuthal  parameters  $K'$ and $K''$ satisfy the condition $K' + K'' = 2
\tau N$ are the same.
As for extreme values  of  gauge  coupling constants, the gauge field $A_{\mu}$
is decoupled from the vortex and Q-ball components as $q= \tau e \rightarrow 0$
and is expressed in terms of the scalar fields $\phi$ and $\chi$ as $q = \tau e
\rightarrow \infty$.
The latter means that the  gauge  field $A_{\mu}$ ceases to be a dynamic object
as $q = \tau e \rightarrow \infty$.

Note that  the  very  possibility   of   the   existence  of  radially  excited
vortex-Q-ball states results from  the  nonlinear  character  of mixed boundary
value problem (\ref{III:3})--(\ref{III:6}) and (\ref{III:8}).
Indeed, a  linear  homogeneous  boundary  value  problem is the Sturm–Liouville
problem.
Solutions with different numbers of nodes correspond to  different  eigenvalues
of the  Sturm–Liouville  problem.
It follows that these solutions satisfy different differential equations because
the eigenvalue is a parameter of differential equation.
In contrast, a nonlinear boundary value problem with fixed parameters  may have
more than one solution, as  it  is  in  the  case  of the vortex-Q-ball system.
In this case,  the  second  and  subsequent  solutions  of  the mixed nonlinear
boundary  value  problem  correspond  to  the  radially  excited  states of the
vortex-Q-ball system.


\begin{acknowledgements}

This  work   was   supported   by   the   Russian   Science  Foundation,  grant
No. 19-11-00005.

\end{acknowledgements}

\end{document}